\documentstyle[11pt,aaspp4]{article}

\def\be{\begin{equation}}
\def\ee{\end{equation}}
\def\go{\mathrel{\raise.3ex\hbox{$>$}\mkern-14mu
             \lower0.6ex\hbox{$\sim$}}}
\def\lo{\mathrel{\raise.3ex\hbox{$<$}\mkern-14mu
             \lower0.6ex\hbox{$\sim$}}}
\def\Porb{P_{\rm orb}}
\def\al{{\alpha}}
\def\vxi{{\vec\xi}}
\def\br{{\bf r}}
\def\bv{{\bf v}}

\def\bC{{\bf C}}
\def\bOmega{{\bf\Omega}}
\def\eps{{\varepsilon}}

\begin{document}

\title{Dynamical Tides in Rotating Binary Stars} 

\author{Dong Lai}
\affil{Theoretical Astrophysics, 130-33, California Institute of
Technology\\
Pasadena, CA 91125\\
E-mail: dong@tapir.caltech.edu}

\begin{abstract}
We study the effect of rotation on the excitation of
internal oscillation modes of a star by the external gravitational
potential of its companion. 
Unlike the nonrotating case, there are difficulties
with the usual mode decomposition for rotating stars because of
the asymmetry between modes propagating in the direction of rotation
and those propagating opposite to it.
For an eccentric binary system, we derive general expressions for 
the energy transfer $\Delta E_s$ and the corresponding 
angular momentum transfer $\Delta J_s$
in a periastron passage when there is no
initial oscillation present in the star. 
Except when a nearly precise orbital resonance occurs 
(i.e., the mode frequency equals multiple of 
the orbital frequency), $\Delta E_s$ is very close to the
steady-state mode energy in the tide in the presence of dissipation. 
It is shown that stellar rotation can change the strength of
dynamical tide significantly. In particular, retrograde rotation 
with respect to the orbit increases the energy transfer by bringing
lower-order g-modes (or f-mode for convective stars), which couple
more strongly to the tidal potential, into closer resonances
with the orbital motion of the companion.

We apply our general formalism to the problems of tidal capture
binary formation and the orbital evolution of PSR J0045-7319/B-star
binary. Stellar rotation changes the critical impact parameter 
for binary capture. Although the enhancement (by retrograde rotation)
in the capture cross section is at most $\sim 20\%$, the 
probability that the captured system survives disruption/merging 
and therefore becomes a binary
can be significantly larger. It is found that in order to 
explain the observed rapid orbital decay of the PSR J0045-7319 binary
system, retrograde rotation in the B-star is required. 
\end{abstract}

\keywords{binaries: close -- pulsars: individual (PSR J0045-7319)
-- stars: neutron -- stars: oscillations -- stars: rotation 
-- hydrodynamics}

\newpage
\section{Introduction}

The standard weak friction theory for tidal interaction in 
binary stars, first introduced by Darwin (1879), and developed
in detail by many authors (e.g., Alexander 1973; Kopal 1978;
Hut 1981), is based on the assumption of static tide in hydrostatic
equilibrium. In the presence of finite dissipation, 
a tidal lag develops between the tidal bulge and the 
external tidal potential, and the resulting tidal torque drives
the spin and orbital evolution. This theory has been applied
successfully to planet-satellite systems (e.g., Goldreich \& Peale
1968), and can also be used to describe binaries of low-mass
(late-type) stars if appropriate turbulent viscosity due to convective
eddies is used (Zahn 1977; Goodman \& Oh 1997). 
The static tide approximation, however, is not
appropriate for many other binary processes, especially those
involving massive (early-type) stars with radiative envelopes,  
where dynamical excitation and radiative damping of low-frequency
g-modes dominate the tidal interaction (Zahn 1977; Goldreich \&
Nicholson 1989). In these early works, the effect of rotation on the
dynamical tide is neglected or introduced phenomenogically 
(but see Nicholson 1979 and Rocca 1989, where some aspects of
rotational effects are explored),
and only circular (or near circular) binary orbits are considered. 
 
In this paper we study dynamical tides in rotating, eccentric
binary stars. Our work is motivated by the recent 
observations of the PSR J0045-7319/B-star binary 
(Kaspi et al.~1996), which contains a radio pulsar 
and a massive rapidly-rotating B-star
in an eccentric $51$ days orbit, with periastron separation of
only $4$ stellar radii. This system therefore constitutes 
an excellent laboratory for studying dynamical tidal interaction. 
In a preliminary analysis (Lai 1996) of the rapid orbital decay of
the binary as revealed by timing observation 
(Kaspi et al.~1996), it was realized that rotation can have 
rather significant effect on the strength of dynamical tides.
In particular, when the B-star has a retrograde
rotation (with magnitude of one half of the maximum value, for
example) with respect to the orbital motion, the dynamical
tidal energy is increased over the nonrotating value by 
two orders of magnitude. Roughly speaking, this
comes about because retrograde rotation shifts the tidal
resonance (where the mode frequency equals the external driving
frequency) from high-order g-modes to lower-order ones,
which couple much more strongly to the tidal potential.  
It is this enhancement of dynamical tide by retrograde rotation
that allows the binary orbit to decay
on a short time scale of $0.5$ Myr, although it has been suggested
(correctly in our opinion) by Kumar and Quataert (1997; hereafter KQ)
that differential rotation is also required in order to keep the mode
damping time relatively short (see \S6 for more discussion).
In this paper we present a full analysis of
dynamical tides in rotating stars, with applications 
to the orbital evolution of the 
PSR J0045-7319/B-star binary. We shall focus on the 
dynamical aspects of the problem, i.e., those aspects 
that are independent of dissipation mechanisms.
We particularly emphasize the difficulty with the usual mode
decomposition because of the asymmetry between modes with opposite 
signs of pattern speed. Such difficulty has led to quantitatively 
misleading results concerning the tidal energy in stars with 
rapid prograde rotation (KQ; see \S 6.3 for discussion). 

Another prominent example in which dynamical tides play an important
role is binary formation from tidal capture. This process was
originally suggested by Fabian, Pringle \& Rees (1975) to explain the
origin of low-mass X-ray binaries in globular clusters:
During a close passage of two stars in an unbound orbit,
the excess orbital energy is transferred through
dynamical tidal excitation to the internal stellar
oscillations, resulting in a bound system. 
Although it has been recognized in recent years that 
primordial binaries play an important role in the dynamics of globular
clusters, and may also be involved (via exchange interaction
with another star) to produce low-mass X-ray binaries and other stellar
exotica (e.g., Hut et al.~1992; Davies 1996), 
tidal capture can still be an efficient means to produce 
tight binaries, thereby providing an energy source to reverse core
collapse. Many theoretical issues related to tidal capture process 
have been studied, including linear and nonlinear energy transfer
(Press and Teukolsky 1977, hereafter
PT; Lee \& Ostriker 1986; Khokhlov, Novikov \& Pethick 1993),
evolution of the binary subsequent
to capture (Kochanek 1992; Novikov, Pethick \& Polnarev 1992;
Mardling 1995) and dissipation of tidally excited modes
(Kumar \& Goodman 1996). However, the possible role 
of stellar rotation on the tidal energy and angular momentum transfer 
has not been considered. Indeed, we find in this paper that
rotation changes the cross-section for tidal capture, and
can significantly increase the probability of forming binaries
through tidal interaction. 

Our paper is organized as follows. 
In \S 2 we present the basic equations describing the tidal excitations
of normal modes in a rotating star and the orbital response 
to the excited oscillations. We then derive in \S 3 
general expressions for energy transfer and angular momentum transfer
in a periastron passage when there is no initial oscillation. 
For an eccentric system, this energy transfer is directly 
proportional to the steady-state mode energy in the presence of
dissipation, and provide a measure of the strength of the dynamical
tide (see \S 6.1). 
In \S 4 we examine the properties of rotating modes that are
relevant for tidal excitation; these are used in specific
calculations presented in \S\S 5-6. 
We apply our general formalism to the tidal capture problem in \S 5 and 
the orbital evolution of PSR J0045-7319/B-star binary in \S 6. 
A brief discussion of related problems is given in \S 7.

\section{Basic Equations}

Consider a rotating star with mass $M$, radius $R$ and spin
$\bOmega_s$, interacting with a companion $M'$. We treat $M'$ as a
point mass, although generalization to the case where both stars have
finite structure is straightforward. For simplicity here we
assume the spin axis is perpendicular to the orbital plane (but
$\Omega_s$ can have both positive and negative signs, corresponding 
to prograde and retrograde rotations with respect to the orbit).
Generalized equations for arbitrary spin-orbit inclination angles are
summarized in Appendix A. 

In the linear regime, the perturbation of the tidal potential 
on $M$ is specified by the Lagrangian displacement
$\vxi(\br,t)$ of a fluid element from its unperturbed
position. In the inertial frame, the equation of
motion can be written in the form:
\footnote{The mode amplitude can be equivalently studied
in the rotating frame. In this case, the equation of motion
is $\partial^2\vxi/\partial t^2+2\bOmega_s\times
(\partial\vxi/\partial t)
+{\bf C}'\cdot\vxi=-\nabla U$, and in Eq.~(\ref{potential})
we should replace $\Phi$ by $(\Phi-\Omega_s t)$. 
Our treatment in the inertial frame is more convenient for
generalizatiion to the case of arbitrary spin-orbit inclination angles
(Appendix A).}
\be
{\partial^2\over\partial t^2}\vxi+2(\bv\cdot\nabla){\partial\vxi
\over\partial t}+\bC\cdot\vxi=-\nabla U, 
\label{xi}\ee
where $\bv=\bOmega_s\times\br$ is the unperturbed fluid
velocity, $\bC$ is a self-adjoint operator (see Lyden-bell \& Ostriker
1967). The external potential in Eq.~(\ref{xi})
is given by
\be
U(\br,t)=-{GM'\over |\br-{\bf D}(t)|}=-GM'\sum_{lm}W_{lm}
{r^l\over D^{l+1}}\exp(-im\Phi)Y_{lm}(\theta,\phi),
\label{potential}\ee
where we have chosen a spherical coordinate system centered
at $M$ with the $z$-axis long the rotation axis and the $x$-axis
pointing to the periastron:
$\br$ is the position vector of a fluid element the star,
and ${\bf D}(t)=[D(t),\pi/2,\Phi(t)]$ 
specifies the position of the point mass $M'$ 
[$D(t)$ is the orbital separation, $\Phi(t)$ is the orbital phase],
and $W_{lm}$ is a coefficient as defined in PT:
\be
W_{lm}=(-)^{(l+m)/2}\left[{4\pi\over 2l+1}(l+m)!(l-m)!\right]^{1/2}
\left[2^l\left({l+m\over 2}\right)!\left({l-m\over
2}\right)!\right]^{-1},
\ee
[here the symbol $(-)^k$ is to be interpreted as zero when $k$ is not
an integer]. In equation (\ref{potential}), the $l=0$ and $l=1$ terms
should be dropped, since they are not relevant to tidal deformation.
For $l=2$, the only nonzero $W_{lm}$'s are 
$W_{2\pm 2}=(3\pi/10)^{1/2}$ and $W_{20}=-(\pi/5)^{1/2}$,
reflecting the symmetry of the tidal potential with
respect to $\phi\rightarrow \phi+\pi$. Thus
only modes with $m=0,~\pm 2$ are excited.

An eigenmode $\vxi_\alpha(\br,t)=\vxi_\alpha(\br)e^{i\sigma_\alpha t}
\propto e^{i\sigma_\alpha t+im\phi}$
satisfies the equation
\be
-\sigma_\alpha^2\vxi_\alpha+2i\sigma_\alpha(\bv\cdot\nabla)\vxi_\alpha
+\bC\cdot\vxi_\alpha=0,
\label{mode}\ee
where $\sigma_\alpha$ is the mode angular frequency in the inertial
frame, and $\alpha=\{njm\}$ specifies the mode index: $n$ gives the
number of nodes in the radial eigenfunction (the order of the mode),
$j$ specifies the number of nodes in the $\theta$-eigenfunction
($j$ reduces to $l$ in the nonrotating limit), and $m$ is the
azimuthal mode index. We shall use the convention 
$\sigma_\al>0$ in this paper, so that a $m>0$ mode has
a retrograde pattern speed ($-\sigma/m<0$), while a
$m<0$ mode has a prograde pattern speed. 
Although the normal modes are not complete in
strict mathematical sense, any reasonable initial data
for Eq.~(\ref{xi}) evolve in time as a linear superposition of
the eigenmodes $\vxi_\alpha(\br)$
(see Dyson \& Schutz 1979 for an analysis of the completeness
problem). Thus we write
$\vxi(\br,t)=\sum_\alpha\!a_\alpha(t)\vxi_\alpha(\br)$,
and normalize the eigenmode via
$\int\!d^3x\,\rho\,\vxi_\alpha\cdot\vxi_\alpha=1$.
Substituting this expansion into Eq.~(\ref{xi}) and using
Eq.~(\ref{mode}), we obtain the dynamical equation for the mode
amplitude:
\be
\ddot a_\alpha+\sigma_\alpha^2a_\alpha
-2iB_\alpha(\dot a_\alpha-i\sigma_\alpha a_\alpha)
+2\gamma_\alpha\dot a_\alpha
=\sum_l {GM'W_{lm}Q_{\alpha l}\over D^{l+1}}e^{-im\Phi},
\label{eqa}\ee
where $Q_{\alpha l}$ is the tidal coupling coefficient (the overlap
integral) defined by 
\be
Q_{\alpha l}=\int\!d^3x\,\rho\,{\vec\xi}_\alpha^\ast\cdot\nabla
(r^lY_{lm})=\int\!d^3x\delta\rho_\alpha^\ast(r^lY_{lm}),
\label{qalpha}\ee  
[here $\delta\rho_\alpha=-\nabla\cdot(\rho\vxi_\alpha)$ is the
Eulerian density perturbation], and the function $B_\alpha$ is given
by 
\be
B_\alpha=i\int\!d^3x\rho\,\vxi_\alpha^\ast\cdot(\bv\cdot\nabla)\vxi_\alpha
=-m\Omega_s+i\int\!d^3x\rho\,\vxi_\alpha^\ast\cdot(\bOmega_s
\times\vxi_\alpha).
\label{balpha}\ee
Note that $\rho\vxi_\alpha^\ast\cdot(\bv\cdot\nabla)\vxi_\alpha
=\nabla\cdot(\rho\bv|\vxi_\alpha|^2)-|\vxi_\alpha|^2\nabla\cdot(\rho\bv)
-\rho\vxi_\alpha\cdot(\bv\cdot\nabla)\vxi_\alpha^\ast$, 
and one can easily show that $B_\alpha$ is real. 
In deriving Eq.~(\ref{eqa}), we have assumed that different 
modes are orthogonal to each other, and have neglected 
the off-diagonal terms $B_{\alpha\beta}\equiv
i\int\!d^3x\,\rho\,\vxi_\beta^\ast\cdot
(\bv\cdot\nabla)\vxi_\alpha$. This {\it approximation} allows us to
decompose $\vxi(\br,t)$ into normal modes and
obtain a simple equation for individual mode amplitude. 
The limitation of this approximation
will be examined in the next section [see
the discussion following Eq.~(\ref{eqa2})].

We have also introduced in Eq.~(\ref{eqa}) the quantity
$\gamma_\alpha$, which is proportional to the mode damping rate.
Multiply Eq.~(\ref{eqa}) by $\dot a_\alpha^\ast$ and take the real 
part, we find:
\be 
{dE_\alpha\over dt}=-2\gamma_\alpha|\dot a_\alpha|^2
+\sum_l{GM'W_{lm}Q_{\alpha l}\over D^{l+1}}
\dot a_\alpha^\ast e^{-im\Phi},
\label{dealpha}\ee
where the mode energy is given by
\footnote{The canonical energy associated with perturbation 
$\vxi(\br,t)$ is generally given by 
$E_c=(1/2)\int\!d^3x\,\rho\,(|\partial\vxi/\partial t|^2
+\vxi^\ast\cdot\bC\cdot\vxi)$. For
$\vxi(\br,t)=a_\alpha(t)\xi_\alpha(\br)$, the canonical energy
reduces to Eq.~(\ref{emode}).}
\be
E_\alpha={1\over 2}|\dot a_\alpha|^2
+{1\over 2}(\sigma_\alpha^2-2B_\alpha\sigma_\alpha)|a_\alpha|^2.
\label{emode}\ee
The last term in Eq.~(\ref{dealpha}) gives the rate at which energy
is transferred to each mode. 
For free oscillation (without external driving), 
$a_\alpha\propto e^{i\sigma_\alpha t}$, the mode energy
becomes $\sigma_\alpha(\sigma_\alpha-B_\alpha)|a_\alpha|^2$, and
Eq.~(\ref{dealpha}) reduces to $dE_\alpha/dt=-2\Gamma_\alpha
E_\alpha$, with the mode amplitude damping rate given by
\be
\Gamma_\alpha={\sigma_\alpha\gamma_\alpha\over\sigma_\alpha-B_\alpha}.
\ee

The excited modes affect the orbital motion through the tidal
interaction potential 
\be
V_{\rm tide}=-GM'\int\!d^3x{\delta\rho(\br,t)\over|{\bf D}-\br|}
=-\sum_{\alpha l}{GM'W_{lm}Q_{\alpha l}\over D^{l+1}}
a_\alpha(t)\,e^{im\Phi}.
\ee
Thus the equations of motion for the orbit are
\begin{eqnarray}
\mu\ddot D &=&\mu D\dot\Phi^2-{MM'\over D^2}-
\sum_{\alpha l}(l+1){GM'W_{lm}Q_{\alpha l}\over D^{l+2}}
e^{im\Phi}a_\alpha(t), \label{eom1}\\
{d(\mu D^2\dot\Phi)\over dt} &=&\sum_{\alpha l}im{GM'W_{lm}Q_{\alpha l}
\over D^{l+1}}e^{im\Phi}a_\alpha(t).
\label{eom2}
\end{eqnarray}

Given the mode properties $(\sigma_\alpha,Q_{\alpha l})$,
equations (\ref{eqa}),(\ref{eom1}) and (\ref{eom2}) completely
determine the time evolution of 
the tidally excited oscillations and the binary orbit.
In general, the energy transfer between
the star and the orbit in a periastron passage depends on the
phases of the oscillation modes and thus varies from one
orbit to another (e.g., Kochanek 1992; Mardling 1995). 
For a highly eccentric system (such as 
parabolic tidal capture and the PSR J0045-7319 binary, to be
discussed in \S 5 and \S 6, respectively), an important measure of the
strength of dynamical tide is the energy transfer to
the star, $\Delta E_s$, and the correspinding angular
momentum transfer, $\Delta J_s$, during 
the ``first'' peraistron when there is no initial
oscillation. General expressions for $\Delta E_s$ and 
$\Delta J_s$ are derived in the next section (\S 3). 
The relation between $\Delta E_s$ and
the steady-state tidal energy $\langle E_s\rangle$ in elliptical 
binary system in the presence of dissipation 
is established in \S 6.1.

\section{Energy Transfer and Angular Momentum transfer}

Here we derive the expressions for the
energy transfer $\Delta E_s$ and angular momentum transfer
$\Delta J_s$ to the oscillation modes in a single periastron passage
assuming the star has no initial oscillation. 
The binary orbit is characterized by the periastron distance $D_p$
and two dimensionless quantities: the eccentricity $e$, and
the ratio between the periastron passage time and the dynamical
time of the star:
\be
\eta=\left({M\over M_t}\right)^{1/2}\left({D_p\over R}\right)^{3/2}
=(1+e)^{1/2}{1\over\hat\Omega_p},
\ee
where $M_t=M+M'$ is the total binary mass, and $\hat\Omega_p$
is the orbital angular 
frequency at periastron in units of $(GM/R^3)^{1/2}$.
We outline two derivations of $\Delta E_s$ and $\Delta J_s$.

Our first derivation is patterned after PT. 
Since for very eccentric system, the orbital period is much longer
than the timescale for periastron passage, and thus
much longer than the periods of the tidally-excited normal modes,
we can consider a continuous Fourier transform, viz.
$a_\alpha(t)=\int\!d\sigma\,e^{i\sigma t}\tilde a_\alpha(\sigma)$.
Equation~(\ref{eqa}) then becomes
\be
(\sigma_\alpha^2-\sigma^2)\tilde a_\alpha
+2B_\alpha(\sigma-\sigma_\alpha)\tilde a_\alpha
+2i\gamma_\alpha\sigma\tilde a_\alpha=\sum_l
{GM'\over D_p^{l+1}}Q_{\alpha l}K_{lm}(\sigma),
\label{atilde}\ee
where $K_{lm}(\sigma)$ depends on the orbital trajectory:
\be
K_{lm}(\sigma)={W_{lm}\over 2\pi}\int_{-\infty}^{\infty}\!dt
\left({D_p\over D}\right)^{l+1}\exp(im\Phi+i\sigma t),
\ee
with the integration centered around the periastron (For elliptical
orbit, the upper limit and lower limit of the integration 
should be replaced by $\Porb/2$ and $-\Porb/2$).
The mode amplitude is then given by
\be
a_\alpha(t)=\sum_{l}\int\!d\sigma\, e^{i\sigma t}
{GM'Q_{\alpha l}K_{lm}(\sigma)\over
D_p^{l+1}[\sigma_\alpha^2-\sigma^2+2B_\alpha(\sigma-\sigma_\alpha)
+2i\gamma_\alpha\sigma]}.
\label{eqvxi}\ee
The energy transfer $\Delta E_s$ to the oscillation modes (in the
inertial frame) is obtained via
\be
\Delta E_s=-\int\!dt\int\!d^3x\rho{\partial\vxi\over\partial
t}\cdot\nabla U^\ast
=\int\!dt\sum_{\alpha l}{GM'W_{lm}Q_{\alpha l}\over D^{l+1}}
e^{im\Phi}\dot a_\alpha(t),
\ee
(only the real part of the expression is meaningful). 
Using Eq.~(\ref{eqvxi}), this becomes
\be
\Delta E_s=\sum_{\alpha ll'}{(GM')^2Q_{\alpha l}
Q_{\alpha l'}W_{lm}W_{l'm}\over 2\pi}
\int\!\!\int\!dt\,dt' {e^{im\Phi(t)+im\Phi(t')}\over 
[D(t)]^{l+1}[D(t')]^{l'+1}}
A_\alpha(t,t'),
\label{eqdele1}\ee
where
\be
A_\alpha(t,t')=\int\!d\sigma{i\sigma e^{i\sigma (t+t')}
\over \sigma_\alpha^2-\sigma^2-2B_\alpha(\sigma_\alpha-\sigma)
+2i\gamma_\alpha\sigma}.
\ee
This integral can be easily evaluated using a countour in the complex
plane, yielding
\be
A_\alpha(t,t')={\pi\over\sigma_\al-B_\al}
\left[\sigma_\al e^{i\sigma_\al (t+t')}+(\sigma_\al-2B_\al)
e^{-i(\sigma-2B_\al)(t+t')}\right],
\ee
for $t+t'>0$, and $A_\al(t,t')=0$ for $t+t'<0$. 
The energy transfer is therefore 
\be
\Delta E_s={GM'^2\over R}\sum_{ll'}
\left({R\over D_p}\right)^{l+l'+2}T_{ll'}(\eta,e),
\label{eqdele2}\ee
with
\be
T_{ll'}(\eta,e)=\pi^2\sum_\alpha
{Q_{\al l}Q_{\al l'}\over\sigma_\al-B_\al}\left[\sigma_\al
K_{lm}(\sigma_\al)K_{l'm}(\sigma_\al)
+(\sigma_\al-2B_\al)K_{l-m}(\sigma_\al-2B_\al)
K_{l'-m}(\sigma_\al-2B_\al)\right].
\label{eqt}\ee
In writing Eq.~(\ref{eqdele2}), we have factored out all dimensional 
physical quantities so that $T_{ll'}$ is dimensionless, i.e., all
quantities in $T_{ll'}$ are in units such that $G=M=R=1$.
The angular momentum transfer is obtained via
\be
\Delta J_s=\int\!dt\int\!d^3x\,\delta\rho\left(-{\partial
U^\ast\over\partial\phi}\right)
=\int\!dt\sum_{\alpha l}(-im){GM'W_{lm}Q_{\alpha l}
\over D^{l+1}}e^{im\Phi}a_\alpha(t).
\label{delj1}\ee
Using a similar procedure we find:
\be
\Delta J_s={GM'^2\over R}\left({R^3\over GM}\right)^{1/2}
\sum_{ll'}\left({R\over D_p}\right)^{l+l'+2}S_{ll'}(\eta,e),
\label{eqdelj2}\ee
with
\be
S_{ll'}(\eta,e)=\pi^2\sum_\alpha
{Q_{\al l}Q_{\al l'}(-m)\over\sigma_\al-B_\al}\left[
K_{lm}(\sigma_\al)K_{l'm}(\sigma_\al)
-K_{l-m}(\sigma_\al-2B_\al)K_{l'-m}(\sigma_\al-2B_\al)\right].
\label{eqs}\ee

For a given $\alpha=\{njm\}$, the first terms inside the square
brackets of Eq.~(\ref{eqt}) and Eq.~(\ref{eqs}) correspond
to a mode with frequency $\sigma_\alpha$, while the
second terms correspond to a mode with frequency
$-(\sigma_\al-2B_\alpha)$. To see this more clearly, consider another
derivation of $\Delta E_s$ and $\Delta J_s$. We simply integrate
Eq.~(\ref{eqa}) directly to obtain $a_\al(t)$:
\be
a_\al(t)=\sum_l{GM'W_{lm}Q_{\al l}\over 2i(\sigma_\al-B_\al)}
\left[e^{i\sigma_\al t}\!\int_{-\infty}^t\!\!\!dt'
\,{e^{-im\Phi(t')-i\sigma_\al t'}\over D^{l+1}(t')}
-e^{-i(\sigma_\al-2B_\al)t}\!\int_{-\infty}^t\!\!\!dt'
\,{e^{-im\Phi(t')+i(\sigma_\al-2B_\al)t'}\over D^{l+1}(t')}\right],
\label{eqa1}\ee
where we have neglected the damping term by setting $\Gamma_\al=0$. 
Suppose the periastron passage occurs at $t=0$, then for $t>>0$, we
have
\be
a_\al(t)=\sum_l{\pi GM'Q_{\al l}\over i D_p^{l+1}(\sigma_\al-B_\al)}
\left[e^{i\sigma_\al t}K_{lm}(\sigma_\al)
-e^{-i(\sigma_\al-2B_\al)t}K_{l-m}(\sigma_\al-2B_\al)\right].
\label{eqa2}\ee
Using Eq.~(\ref{emode}) we find that the total mode energy 
after the periastron passage is again given by
Eqs.~(\ref{eqdele2})-(\ref{eqt}).

Equation (\ref{eqa2}) clearly reveals that after excitation 
in a periastron passage, a mode $\vxi(r,\theta)e^{im\phi}$
oscillates at frequencies $\sigma_\alpha$ (its eigenfrequency)
and $-(\sigma_\al-2B_\al)$. How can a mode oscillate at
a frequency other than its own eigenfrequency? 
This conundrum comes from our approximate treatment of
mode decomposition leading to Eq.~(\ref{eqa}). 
Note that the mode $\vxi_\alpha\propto e^{-i\sigma t-im\phi}$
is physically identical to the mode $\vxi_\al\propto
e^{i\sigma+im\phi}$.
For a given $n,j$ and $|m|=2$ (for example), there are 
four mathematically distinct modes:
\footnote{It can be shown that if $\vxi(r,\theta)e^{im\phi+i\sigma t}$
is a solution of Eq.~(\ref{mode}), then $\vxi\,'(r,\theta)
e^{-im\phi-i\sigma t}$ is also a solution, where
$\vxi\,'=(\xi_r,\xi_\theta,-\xi_\phi)$. These two modes have the same
$\delta\rho(r,\theta)$. See Schutz (1979).}
\begin{eqnarray}
\vxi_m(\br,t)&=& \vxi_2(r,\theta)e^{i2\phi+i\sigma_2t},~~~~~~
\vxi_2^{~'}(r,\theta)e^{-i2\phi-i\sigma_2t},~~~~~~({\rm
retrograde~mode});\label{modes1}\\
\vxi_m(\br,t)&=& \vxi_{-2}(r,\theta)e^{-i2\phi+i\sigma_{-2}t},~
\vxi_{-2}^{~'}(r,\theta)e^{i2\phi-i\sigma_{-2}t},
~~~~~({\rm prograde~mode}),
\label{modes}\end{eqnarray}
and only two of them are physically distinct, one with prograde
pattern speed and another with retrograde pattern speed.
Clearly, within the context of our approximation, we
should identify $(\sigma_m-2B_m)$ with $\sigma_{-m}$ (where the
other mode indices $nj$ are surpressed).
This $(-\sigma_{-m})$ mode has wavefunction $\vxi_{-m}^{~'}(r,\theta)$,
different from $\vxi_m(r,\theta)$, and thus its ampitude should be
proportional to $Q_{-m}$ rather than $Q_m$ as indicated in
Eq.~(\ref{eqa2}). But this is an artifact of our
approximation. In the slow rotation limit ($|\Omega_s|<<\sigma_\al$), 
the mode frequencies are related to the nonrotating value $\sigma^{(0)}$
by $\sigma_m=\sigma^{(0)}+B_m$ and $\sigma_{-m}=\sigma^{(0)}+B_{-m}
=\sigma^{(0)}-B_m$, while the wavefunctions are unchanged to 
the leading order (so that $Q_m=Q_{-m}=Q^{(0)}$).
Thus Eqs.~(\ref{eqdele2}),~(\ref{eqt}),~(\ref{eqdelj2}),~(\ref{eqs})
and (\ref{eqa2}) are exact in this limit. For larger rotation rates,
they should still be good approximations as long as $\sigma_m-2B_m$
is close to $\sigma_{-m}$, and the difference between $Q_m$ and
$Q_{-m}$ is small --- We find that these are satisfied for the
relevant modes considered in \S5 and \S6.

The preceding discussion leads us to adopt the following 
{\it ansatz}: The amplitude of each mode (specified by 
$\al=\{njm\}$ {\it and} the direction of pattern speed) is given by 
the first term of Eq.~(\ref{eqa1}), while the contribution of the 
second term to $\Delta J_s$ and $\Delta E_s$ is identical to 
that of a corresponding $(-m)$ mode with the same pattern speed. 
The dimensionless energy trnasfer $T$ and angular momentum
transfer $S$ are therefore given by
\begin{eqnarray}
T_{ll'}(\eta,e)&=&2\pi^2\sum_\alpha {Q_{\al l}Q_{\al l'}
\over\sigma_\al-B_\al}\sigma_\al
K_{lm}(\sigma_\al)K_{l'm}(\sigma_\al),\label{eqt2}\\
S_{ll'}(\eta,e)&=&2\pi^2\sum_\alpha
{Q_{\al l}Q_{\al l'}(-m)\over\sigma_\al-B_\al}
K_{lm}(\sigma_\al)K_{l'm}(\sigma_\al),\label{eqs2},
\end{eqnarray}
where the factor of $2$ results from the equivalence of 
the $(m,\sigma_\al)$ mode and the $(-m,-\sigma_\al)$ mode.
Since the mode pattern speed is
$(-\sigma_\alpha/m)$, the contribution of each mode to the energy
transfer $\Delta E_\alpha$ and angular momentum transfer 
$\Delta J_\alpha$ are related by $\Delta J_\alpha=(-m/\sigma_\alpha)
\Delta E_\alpha$, as seen from Eqs.~(\ref{eqt2}) and (\ref{eqs2}).
It is also useful to consider the total mode energy
$\Delta E_s^{(r)}$ in the star's rotating frame. 
Since the rotational kinetic energy associated with the
angular momentum transfer is $\Omega_s\Delta J_s$, we have
$\Delta E_s^{(r)}=\Delta E_s-\Omega_s\Delta J_s$, and the
corresponding dimensionless energy is 
\be
T_{ll'}^{(r)}(\eta,e)=T_{ll'}(\eta,e)-\Omega_s S_{ll'}(\eta,e)
=2\pi^2\sum_\alpha {Q_{\al l}Q_{\al l'}
\over\sigma_\al-B_\al}\omega_\al
K_{lm}(\sigma_\al)K_{l'm}(\sigma_\al),\label{eqtr}\\
\ee
where $\omega_\al=\sigma_\al+m\Omega_s$ is the mode frequency in 
the rotating frame.
We emphasize that, although our procedure leading to 
Eqs.~(\ref{eqt2})-(\ref{eqtr}) is somewhat ad-hoc
mathematically [The ``rigorous'' derivation of
Eqs.~(\ref{eqt}),~(\ref{eqs}) and (\ref{eqa2}) are exact only in the
slow-rotation limit], the forms of these expressions are 
naturally expected on the physical ground:
the retrograde mode should be
related to $|Q_2K_{l2}(\sigma_2)|^2$ and the prograde mode to
$|Q_{-2}K_{l-2}(\sigma_{-2})|^2$. The only uncertainty 
is the factor $(\sigma_\al-B_\al)/\omega_\al=1-(i/\omega_\al)
\int\!d^3x\,\rho\vxi_\al^\ast\cdot(\bOmega_s\times\vxi_\al)$
[cf.~Eq.~(\ref{eqtr})] which only serves as a
insignificant ``correction'' to $Q_\al$ (see \S6 for further
discussion).

Although the above expressions apply for general $l$, 
in this paper we shall consider only the quadrupolar ($l=2$)
tides. The contribution of higher-$l$ tidal potential to $\Delta E_s$
is of order $(R/D_p)^2$ smaller than that of the leading $l=2$ term. 

\section{Properties of Modes in Rotating Stars}

We consider two simple stellar models in this paper: 
(i) A $\Gamma=\Gamma_1=5/3$ polytrope ($\Gamma$ is the
polytropic index, $\Gamma_1$ is the adiabatic index), representing 
low-mass ($M\lo 0.5M_\odot$) MS star with thick
convective envelope; (ii) A $\Gamma=4/3,~\Gamma_1=5/3$ 
polytrope, representing high-mass (larger than a few solar mass)
star for which the envelope is radiative; it is also 
a good approximation to the outer structure of intermediate-mass stars
($\sim 1M_\odot$). 

We examine the effects of rotation on the mode properties (frequency
and tidal coupling coefficient) relevant to tidal excitation. 

\subsection{f-Mode and p-Modes}

For low-mass, convective star (which does not have any g-mode), 
the f-mode and lowest order p-modes are the most important 
modes that absorb tidal energy. For
$\Omega_s$ much smaller than the mode frequency, 
perturbation theory is adequate (e.g., Unno et al.~1989).  
To leading order, the mode wavefunction is unchanged, 
and the mode frequency is modified from the nonrotating value
$\sigma_{nl}^{(0)}$ by
\be
\sigma_{nlm}=\sigma_{nl}^{(0)}+B_{nlm}
=\sigma_{nl}^{(0)}-m\Omega_s+mC_{nl}\Omega_s,
\ee
where $m\Omega_sC_{nl}=i\int\!d^3x\,\rho\,\vxi_\al^\ast\cdot 
(\bOmega_s\times\vxi_\al)$. 
Using the zeroth order eigenfunction 
$\vxi_{nlm}(\br)=\left[\xi_r(r){\bf e}_r+\xi_\perp(r)
{\hat\nabla}_\perp\right]Y_{lm}(\theta,\phi)$, where 
$\hat\nabla={\bf e}_\theta(\partial/\partial\theta)
+({\bf e}_\phi/\sin\theta)(\partial/\partial\phi)$,
we have
\be
C_{nl}=\int_0^R\rho r^2(2\xi_r\xi_\perp+\xi_\perp^2)\,dr,
\label{cnl}\ee
and the eigenfunction is normalized via
$\int\!d^3x\rho\,\vxi^\ast\cdot\vxi=\int\!dr\,\rho\,r^2[\xi_r^2+l(l+1)
\xi_\perp^2]=1$.
In the rotating frame the mode angular frequency is 
$\omega_{nlm}=\omega_{nl}^{(0)}+mC_{nl}\Omega_s$.
For the $\Gamma=5/3$ polytrope, we have
$C_{02}=0.4955$ (f mode), $C_{12}=0.1525$ (p$_1$ mode), and
$C_{22}=0.0787$ (p$_2$ mode); for the
$\Gamma=4/3,\,\Gamma_1=5/3$ polytrope, we have 
$C_{02}=0.2543$ (f mode), $C_{12}=0.1538$ (p$_1$ mode), and
$C_{22}=0.0818$ (p$_2$ mode).
For higher $\Omega_s$, linear perturbation theory breaks down, but
accuarte numerical results are available (e.g., Managan 1986; 
Ipser \& Lindblom 1990. The later is essentially an exact method).
In our calculation of tidal capture below (\S 5), we shall adopt the
linear approximation at small $\Omega_s$ with correction for high
$\Omega_s$ using the results given in Managan (1986). 
The modification to the mode eigenfunction is negligible
and therefore we set $Q_\al=Q_{\alpha 2}$ to be the same as that of a
nonrotating star. 

\subsection{g-Modes}

For intermediate to massive stars with radiative envelopes, g-modes
dominate the tidal energy transfer because their relatively low 
frequencies match the orbital frequency at periastron (see \S\S 5-6). 
Since we shall consider the cases where the rotation frequency 
is comparable or larger than the g-mode frequency, 
the perturbation theory is not adequate. 
An approximate treatment of g-modes in rapidly rotating star 
is based on the so-called ``traditional approximation''
(Chapman \& Lindzen 1970; Unno et al.~1989;
Bildsten et al.~1996), where the centrifugal
distortion of the star is
neglected, as well as the Coriolis forces associated with the
horizontal component of the spin angular velocity. Note that 
this approximation is strictly valid only for high-order g-modes for
which $\omega_\alpha<<N$ (where $N$ is the 
Brunt-V\"ais\"al\"a frequency) and $\xi_\perp>>\xi_r$, and for
$\Omega_s<<N$ (so that the radial component of the Coriolis force can be 
neglected compared to the buoyancy force), but it also provides
a good estimate even for low-order modes.

Neglecting the perturbation in the gravitational potential (Cowling
approximation) and adopting the traditional
approximation, the radial Lagrangian displacement and 
Eulerian pressure perturbation can be written as
\be
\xi_r(\br)=\xi_r(r)H_{jm}(\theta)e^{im\phi},~~~
\delta P(\br)=\delta P(r)H_{jm}(\theta)e^{im\phi},
\label{eigen}\ee
where $H_{jm}(\theta)$ is the Hough function, satisfying the 
Laplace tidal equation (Chapman \& Lindzen 1970).
Separating out the angular dependence, the fluid continuity equation
and Euler equation (in the rotating frame) 
reduce to a set of coupled radial equations:
\begin{eqnarray}
{d\over dr}\delta P &=&-{g\over c_s^2}\delta
P+\rho(\omega^2-N^2)\xi_r,\label{eqdelp}\\
{d\over dr}(r^2\xi_r) &=& {g\over c_s^2}(r^2\xi_r)
-{r^2\over c_s^2\rho}\delta P+{\lambda\over\omega^2\rho}\delta P,
\label{eqxir}
\end{eqnarray}
where the mode index $\alpha=\{njm\}$ has been surpressed, and
$g>0$ is the gravitational acceleration, $c_s$ is the sound speed.
The eigenvalue of the Laplace tidal equation, $\lambda$, depends
on $m$ and the ratio $q=2\Omega_s/\omega_\alpha$. For $q\rightarrow
0$ (nonrotating case), the function $H_{jm}(\theta)e^{im\phi}$
becomes $Y_{lm}(\theta,\phi)$ while $\lambda$ degenerates into
$l(l+1)$.

The properties of the Hough function have been extensively studied
(Longuet-Higgins 1967). We have adopted
the numerical approach of Bildsten et al.~(1996)
in our calculation. For given $q$ and $m$, the eigenvalue
$\lambda$ is obtained by solving the angular equation. The radial
equations (\ref{eqdelp})-(\ref{eqxir}) are then solved together with
appropriate boundary conditions to obtain $\omega$, from which the
actual rotation rate $\Omega_s=q\omega/2$ is recovered. Figure 1 shows
the frequencies (in the rotating frame)
of several $j=2$ g-modes against the rotation rate (Both 
$\omega$ and $\Omega_s$ are plotted in units of $\omega^{(0)}$, the 
mode frequency at zero rotation). These modes have $l=2$ in the 
$\Omega_s=0$ limit. For high order g-modes, a WKB analysis for the
radial equations gives $\omega\propto\sqrt{\lambda}$. Thus we have 
$\omega/\omega^{(0)}=\sqrt{\lambda/6}$ and
$\Omega_s/\omega^{(0)}=q\sqrt{\lambda/24}$. 
For $|\Omega_s/\omega^{(0)}|<1.1$ (which is 
satisfied by g-modes with $n\le 13$ when $\Omega_s<0.5$),
our numerical results for the g-modes of the $\Gamma=4/3$,
$\Gamma_1=5/3$ polytrope can be fitted by the following analytic
expressions to within $1\%$: 
\footnote{ 
In these expressions, the linear and quadratic terms
are derived from an analytic expansion of $\lambda$ for
small $q$ (Bildsten \& Ushomirsky 1996, private communication), 
and the last terms 
are based on numerical fitting.}
\begin{eqnarray}
\bar\omega &=& 1+\left[(1/3)\bar\Omega_s+(13/42)\bar\Omega_s^2
-0.064\bar\Omega_s^{4.6}\right]\,f,  ~~~~~~(m=2),\label{eqm2}\\
\bar\omega &=&
1+\left[(6/7)\bar\Omega_s^2-0.31 |\bar\Omega_s|^{3.3}\right]\,f,
~~~~~~~~~~~~~~~~~~~~~~(m=0),\label{eqm0}\\
\bar\omega &=& 1+\left[-(1/3)\bar\Omega_s+(13/42)\bar\Omega_s^2
-0.118\bar\Omega_s^3\right]\,f,  ~~~~~~(m=-2),\label{eqm-2}
\end{eqnarray}
where $\bar\omega\equiv\omega/\omega^{(0)}$, $\bar\Omega_s\equiv
\Omega_s/\omega^{(0)}$, and the factor $f\le 1$ 
depends on specific g-modes: In the 
WKB limit (high-order g-modes with $n\rightarrow\infty$) 
we have $f=1$. The values of $f$ for other selected modes are:
$f=0.48,~0.81,~0.9,~0.95$ for $g_1,~g_5,~g_{10},~g_{20}$.
Equations (\ref{eqm2})-(\ref{eqm-2}) break down when
$|\bar\Omega_s|\go 1.1$. In this high-$\bar\Omega_s$ regime, 
the following asymptotic expressions can be applied:
$\bar\omega\simeq 1.54\sqrt{\bar\Omega_s}$ (for $m=0,~2$), and
$\bar\omega\simeq 0.82$ (for $m=-2$). 
Note that these expressions [Eqs.~(\ref{eqm2})-(\ref{eqm-2})] are
valid only for $\Omega_s\ge 0$. The eigenfrequencies for 
$\Omega_s<0$ can be obtained using the relation
$\omega_m(\Omega_s)=\omega_{-m}(-\Omega_s)$.

Rotation also changes the tidal coupling coefficient $Q_{\alpha 2}$,
although the correction is not significant since all the g-modes 
we include in our calculation ($n\le 20$) that are strongly excited 
satisfy $\Omega_s/\omega^{(0)}<1.7$, i.e., $\Omega_s$ is not much larger
than the mode frequency. 
Expressions for evaluating the $Q_{\alpha l}$ and $B_\alpha$ are given 
in Appendix B. Figure 2 shows the coupling coefficient
$Q_{\alpha}=Q_{\al 2}$ for several g-modes.  

For rotating stars, the $j=4,~6,~8,\cdots$ (corresponding to
$l=4,~6,~8,\cdots$ in the $\Omega_s\rightarrow 0$ limit)
modes with $m=0,\pm 2$ are also coupled to the quardrupolar
tidal potential. (The coupling with the $j=3,~5,~7,\cdots$ modes 
vanishes because the eigenfunctions are odd with respect to 
the equator while the tidal potential is even). 
The coupling goes to zero at $\Omega_s=0$. Some properties of the
$j=4$ modes are plotted in Figure 3 and compared with the
$j=2$ modes. Two factors make these higher-$j$ modes less important
to energy transfer during a tidal encounter than the $j=2$ modes:
(i) The $j\ge 4$ modes have higher frequencies than the
$j=2$ modes of the same radial order, therefore resonance
condition (see next section) is satisfied for higher-order modes,
which couple less strongly to the tidal potential;
(ii) The $j\ge 4$ modes have much smaller coupling coefficients
than the $j=2$ modes of the same radial order (The angular integral
$Q_\theta$ is plotted in Fig.~3(b); the radial integral
$Q_r$ is slightly smaller for the $j=4$ modes than for the $j=2$
modes). We have checked that the contribution of the $j>2$
modes to the energy transfer is always less than a few percent. 
Thus they are neglected in our calculations below.  

\subsection{r-Modes}

For nonrotating stars, these are ``trivial'' toriodal modes
with zero frequency and zero density perturbation, 
and thus they do not couple to the tidal potential.
With rotation these modes are mixed with 
spheroidal components, and attain finite frequencies.
The r-mode (sometimes also called ``quasi-toroidal mode'')
is the generalized form of Rossby waves on a 
spherical surface as studied in terrestrial meteorology. 
For small rotation rate $|\Omega_s|<<1$, r-modes can be
studied using perturbation expansion in powers of $\Omega_s$
(Provost et al.~1981; Saio 1982). 
To linear order in $\Omega_s$, the mode frequency increases
as $\omega=2m\Omega_s/l(l+1)$, while its Lagrangian displacement
remains purely toroidal. Finite density perturbation comes
in only in the second order of $\Omega_s$. 
For higher $\Omega_s$, r-modes can also be associated with eigenvalues of
the Laplace tidal equation (Longuet-Higgins 1967). 
Since the coupling between r-modes and the tidal potential is weak
($Q_\al\propto\Omega_s^2$),
we shall neglect them in our calculations below.  

\section{Tidal Capture Binary Formation}

We now apply the results of \S\S3-4 to study the effect
of rotation on tidal capture.
As long as the relative velocity between the stars
at infinite separation is much less than the escape velocity from
the star, a parabolic orbit is a good approximation.
We consider two representative stellar models in \S 5.1 and \S 5.2.

\subsection{Intermediate to Massive Stars: $\Gamma=4/3$ Polytrope}

For the $\Gamma=4/3,\,\Gamma_1=5/3$ polytrope, the functions
$T_2^{(r)}=T_{22}^{(r)}$,~$S_2=S_{22}$ 
and $T_2=T_2^{(r)}+\Omega_sS_2$ [where
$\Omega_s$ is the rotation rate in units of $(GM/R^3)^{1/2}$]
are plotted against $\eta$ in Figure 4 for $\Omega_s=0,\,\pm 0.2$ 
and $\pm 0.4$. We clearly see that for $\Omega_s<0$ (retrograde
rotation), the energy and angular momentum transfers can be
significantly increased over the nonrotating values. 
This comes about for the following reasons:
During a perastron passage, the most strongly
excited modes are those (i) propagating in the same direction
as the orbital motion of the companion (corresponding to
the $m=-2$ modes in our notation), (ii) 
having frequencies in the inertial frame
comparable to the ``driving frequency'', which is equal to
twice of orbital frequency at periastron, i.e.,
$\sigma_\alpha=\omega_\alpha-m\Omega_s=2\Omega_p$ (the ``resonant
condition''), and (iii) having relatively large $Q_\alpha$.
Since higher-order (lower frequency) g-modes have smaller
couping coefficients than the low-order ones, the trade-off
between (ii) and (iii) implies that the modes that absorb
most tidal energy are those with frequencies higher
than the resonant mode. For example, when $\eta=7$ and $\Omega_s=0$,
the resonant mode is $g_{15}$, with frequency $\omega_\alpha\simeq 0.4
\simeq 2\Omega_p$, while the dominant modes in energy transfer 
are g$_6$-g$_8$ (which have $\omega_\al=0.85-0.68$).
A retrograde rotation ``drags'' such $m=-2$
waves backward, so that the mode frequencies in the inertial frame
are lowered. As a result, 
the dominant modes in energy transfer 
are shifted to lower radial orders.
Because $Q_\alpha$ increases rapidly with decreasing mode order
or increasing mode frequency (we find from numerical calculations 
that $Q_\alpha\propto \omega_\al^{4.5}$ for nonrotating $\Gamma=4/3$
polytrope), the energy transfer is greatly increased.
For example, at $\Omega_s=-0.4$ and $\eta=7$, the dominant
modes are g$_3$-g$_5$, the net energy transfer (in the
inertial frame) $\Delta E_s$ is about $20$ times larger than the
nonrotating value. 

While relatively small prograde rotation ($\Omega_s>0$)
decreases the energy transfer, we see from Fig.~4 that
for sufficiently  large $\Omega_s$ (or more precisely for 
sufficiently large ratio of $\Omega_s/\Omega_p$), 
$\Delta J_s$ or even $\Delta E_s$ can become
negative, i.e., angular momentum and energy are
transferred from the star to the orbit (see also Fig.~8 in \S 6).
This behavior can be understood as follows:
The angular momentum of a mode is related to its energy in the
rotating frame by $J_\al=(-m/\omega_\al)E_\al^{(r)}$, 
thus the prograde mode ($m=-2$) carries positive angular momentum,
while the retrograde mode carries negative angular momentum
(Note that $E_\al^{(r)}$ is always positive).  
For a fixed $\eta$, prograde rotation increases the frequencies of
the prograde modes, shifts the resonance to higher radial orders
(which have smaller $Q_\alpha$), therefore decreases the
positive angular momentum transfer to the star.
On the other hand, the retrograde modes are ``dragged''
forward by the rotation, and their $\sigma$'s
become smaller and may even become negative (i.e.,
they are prograde in the inertial frame). 
Therefore their contributions to the energy transfer become
increasingly more important and the negative angular momentum 
transfer to the star increases as $\Omega_s$ increases.  
When the negative angular momentum transferred to the $m=2$ modes
becomes larger than the positive angular momentum transferred
to the $m=-2$ modes, the net $\Delta J_s$ changes sign,
and at the same time, $\Delta E_s^{(r)}$ begins to increase
with increasing $\Omega_s$ (for a given $\eta$). At even
larger $\Omega_s$, when the change in the rotational energy
$\Omega_s\Delta J_s<0$ (due to angular momentum lost 
to the orbit) dominates over the kinetic energy in the modes,
the net $\Delta E_s$ becomes negative.
Our numerical results for the critical $\Omega_s$ at which $\Delta J_s$
or $\Delta E_s$ changes sign can be fitted nicely by
(for $0.1<\Omega_p<0.35$)
\begin{eqnarray}
\Omega_s&=&1.2\Omega_p+0.08,~~~~{\rm for}~\Delta J_s=0\label{dej0}\\
\Omega_s&=&1.5\Omega_p+0.07,~~~~{\rm for}~\Delta E_s=0.\label{de0}
\end{eqnarray}
Kumar \& Quataert (1997) claimed that $\Delta J_s$ and $\Delta E_s$ 
become negative when $\Omega_s\go \Omega_p$. This is 
different from our Eqs.~(\ref{dej0})-(\ref{de0}). 
We believe this quantitative difference arises from their 
problematic treatment of
mode decomposition. We defer a discussion of this point
to the next section (\S 6.3). 

The maximum $D_p$ for capture is determined
by the condition $\Delta E^{(i)}(M)+\Delta E^{(i)}(M')=
\mu v^2/2$, where $\Delta E^{(i)}(M)=\Delta E^{(i)}$ and 
$\Delta E^{(i)}(M')$
are the energy transfer to $M$ and $M'$ respectively, 
$\mu$ is the reduced mass and $v$ is the relative velocity 
of the two stars at infinite separation. For 
$1.8\le\eta\le 3.5$ (the relevant parameter regime for capture)
the function $T_2$ can be fitted to the form
\be
T_2=T_2^{(r)}+\Omega_sS_2=A\,\eta^{-\alpha},
\ee
where the fitting parameters $A,~\alpha$ for several $\Omega_s$
are listed in Table 1. Let $q=M'/M$ and $\lambda=1+\Delta E^{(i)}(M')
/\Delta E^{(i)}(M)$, we then obtain the critical periastron distance
for capture:
\be
D_{\rm cap}=
R\,\left[2\,\lambda\,A\,q\,(1+q)^{(2+\alpha)/2}\right]^{2/3(4+\alpha)}
\left[{v\over (GM/R)^{1/2}}\right]^{-4/3(4+\alpha)}.
\label{dcap}\ee
Figure 5 depicts numerical results for the encounter between
a $1 M_\odot$ star (modeled as a $\Gamma=4/3,~\Gamma_1=5/3$ polytrope)
and a $M'=1.4M_\odot$ point mass (neutron star).
With $\lambda=1$, Equation~(\ref{dcap}) and Table 1 give 
$D_{\rm cap}/R=2.42\,v_{10}^{-0.188}$ for $\Omega_s=0$
[This agrees with Lee \& Ostriker (1986)], 
$D_{\rm cap}/R=2.10\,v_{10}^{-0.136}$ for $\Omega_s=0.5$,
and $D_{\rm cap}/R=2.77\,v_{10}^{-0.230}$ for $\Omega_s=-0.5$,
where $v_{10}\equiv v/(10\,{\rm km}\,{\rm s}^{-1})$.

\subsection{Low Mass stars: $\Gamma=5/3$ Polytrope}

This is more applicable for low-mass (convective) stars
(e.g., those in globular clusters). 
In this case, only the f-mode ($\omega=1.456$ for $\Omega_s=0$)
is strongly excited, while the contribution of p-modes
to the energy transfer is always less than a few percent
because of their larger frequencies and smaller $Q_\alpha$ (e.g.,
for $\Omega_s=0$, the p$_1$ mode has $\omega=3.21$, implying
that it is almost always out of resonance, and its 
$Q_\alpha$ is an order of
magnitide smaller than that of the f-mode). 
We find similar rotational effect on energy transfer as in \S 5.1,
although it is less prominent (since there is only one 
radial mode involved).
For sufficiently large $\Omega_s$, the angular momentum transfer
$\Delta J_s$ changes signs, but this occurs
only at $\Omega_s/\Omega_p\go 3.5$; the energy transfer $\Delta E_s$
never changes sign as long as we restrict to the regime 
$\Omega_s<0.6$ (i.e., $\Omega_s$ is less than the maximum rotation
rate).  

Motivated by the functional form of $K_{lm}$
in the asymptotic limit ($\eta\omega>>1$; see Appendix C), we can fit
the function $T_2$ to the form:
\be
T_2=T_2^{(r)}+\Omega_s S_2=A\,\eta^5\,\exp(-\alpha\eta),
\ee
with the fitting parameters $A$ and $\alpha$ given in Table 1. 
This fitting is valid for the range of $\eta$
near tidal capture.
In the asymptotic limit, including only the $m=-2$ f-mode contribution,
we have $\alpha=4\sqrt{2}\sigma/3$. For
$\Omega_s=0$, this gives $\alpha=2.74$. 
Figure 6 shows numerical results for encounter between
a $1 M_\odot$ star (modeled as a $\Gamma=5/3$ polytrope)
and a $M'=1.4M_\odot$ point mass (neutron star).
Again, we see that the critical capture radius increases
with increasing retrograde rotation. 

\subsection{Discussion on Tidal Capture Binary Formation}

The results of \S\S 5.1-5.2 indicate that stellar rotation can 
increase or decrease the critical capture radius by up to $\sim 20\%$
(for nearly maximumly rotating stars), leading to
similar change in the capture cross-section
$\sigma_{\rm cap}
\simeq {2\pi G(M+M')D_p/v^2}$ (where $v$ is the 
relative velocity of the stars in infinite separation). 
In globular clusters, the stellar rotation is likely to be
much less than the maximum since the stars persumably have slowed
down significantly by magnetic breaking. 
For open clusters (with much lower
velocity dispersion, $\sim 1$ km~s$^{-1}$) or other
young star clusters, such rapid rotation is certainly realistic
(e.g., the typical surface rotation velocity for B stars
is $\sim 400$ km/s, corresponding to a rotation rate close to the
maximum). 
While the $20\%$ change in $D_{\rm cap}$ may seem insignificant
by itself, we note
that the enhancement to the probability for tidal capture binary
formation can be much larger. Indeed, it has been 
suggested that almost all tidally captured bound systems may 
merge during the subsequent periastron passages, leaving no 
binaries behind (e.g., McMillan et al.~1987; 
Kochanek 1992). The critical periastron 
distance for merging, $D_{\rm merge}$, is uncertain, 
but a reasonable lower limit
is given by the sum of the radii of the two stars. Clearly, 
if $D_{\rm merge}$ is close to $D_{\rm cap}$, then
a relatively small increase in $D_{\rm cap}$ can lead to much 
larger increase in the efficiency for binary formation.
If so, we can predict that {\it binaries formed 
by tidal capture must perferentially have retrograde rotations.} 
Although subsequent tidal evolution tends to align the
spin and the orbital angular momentum, the retrograde
signature may still be perserved for relatively young binaries.

\section{Orbit and Spin Evolution of PSR J0045-7319/B-star Binary}

The PSR J0045-7319 binary, containing a $0.93$\,s radio pulsar
and a massive B-star companion ($M=8.8M_\odot,~R=6.4R_\odot$)
in an eccentric ($e=0.808$) 51.17 
days orbit (Kaspi et al.~1994), is unique and important 
in that it is one of the two binary pulsars discovered so far that
have massive main-sequence star companions (The other one is 
PSR B1259-63 with a Be-star companion; Johnston et al.~1994). 
These systems evolve from MS-MS binaries when one of
the stars explode in a supernova to form a neutron star. 
Thus the characteristics of such pulsar binaries can potentially
be used to infer the physical conditions of neutron star
at its formation. The PSR J0045-7319 system, in particular, 
owing to its relatively small orbit and ``clean'' environment,
exhibits two interesting dynamical orbital behaviors:
(i) spin-orbit precessions due
to the rapid, misaligned rotation of the B-star, which strongly
suggests that the neutron star received a kick at birth from
asymmetric supernova (Lai et al.~1995; Kaspi et al.~1996);
(ii) Rapid orbital decay, on a timescale of 
$P_{\rm orb}/\dot P_{\rm orb}=-0.5$ Myr (shorter than the
lifetime of the $8.8M_\odot$ B-star and the characteristic age 
of the pulsar) (Kaspi et al.~1996), which, together with a
generic model of tidal evolution, can be used to constrain the age
of the binary since the supernova and the initial spin of the pulsar
(Lai 1996).  

The mechanism for the orbital decay was discussed in Lai (1996). 
Since mass loss from the B-star is negligible
(as inferred from dispersion measure variation; Kaspi et al.~1996b),
the orbital decay must have a dynamical origin.
The static tide, corresponding to the global, quadrupolar
distortion of the star, cannot explain such rapid orbital decay
because (i) it would require too short a viscous time (less than $30$
days based on our numerical integrations of the tidal equations given by 
Alexander 1973) --- Such a small viscous time
is almost certainly impossible (e.g., even if the star were completely
convective, the viscous time would still be longer than one year);
(ii) It would lead to rapid spin-orbit synchronization 
and alignment even if viscosity were large enough to explain 
the observed $\dot P_{\rm orb}$, in contradiction with observations.
It was suggested that dynamical tidal interaction between the pulsar
and the B-star can explain the rapid orbital decay
provided that the B-star has significant retrograde rotation
with respect to the orbital motion so that the energy transfer
is sufficiently large. The reason that retrograde rotation 
significantly increases the tidal strength is the same as
discussed in \S 5.1, i.e., retrograde rotation shifts
the resonance from high-order g-modes to lower-order ones,
which couple more strongly to the tidal potential.
Several simplifying assumptions were adopted in the preliminary analysis
of Lai (1996). The most important one is that the mode damping times
are assumed to be constant since only small numbers of modes are
strongly excited. 
This assumption is likely to be inadequate, as correctly pointed 
out by KQ, who suggested that an additional
ingredient, differential rotation, is needed to compensate for
the longer damping times of lower-order g-modes. Our focus here,
however, is still to explore the dyanmcal aspects of the tides more
completely. 


\subsection{Dynamical Tidal Energy: Steady State}

The dynamical equations presented in \S 2 for the coupled orbital and
mode evolution can be integrated numerically. Figure 7 depicts
a typical example of such numerical integration. We start the
calculation at the apastron ($t_o=-P_{\rm orb}/2$)
with no initial oscillation, and
include g-modes of order $2-8$ (each has $m=0,\pm 2$ components). 
The total angular momentum $J_{\rm trans}$ and energy $E_{\rm trans}$ 
transferred to the modes are calculated via 
\begin{eqnarray}
J_{\rm trans}(t)&=&\int_{t_o}^t\!dt
\sum_\alpha{GM'W_{lm}Q_\alpha\over D^{l+1}}
(-im)\,a_\alpha e^{im\Phi},\\
E_{\rm trans}(t)&=&\int_{t_o}^t\!dt
\sum_\alpha{GM'W_{lm}Q_\alpha\over D^{l+1}}\,
{\dot a_\alpha}\,e^{im\Phi}.
\end{eqnarray}
Obviously $J_{\rm trans}$ is equal to the mode angular momentum $J_s$,
while $E_{\rm trans}$ is equal to the mode energy
$E_s=\sum_\alpha(1/2)\left[|\dot a_\alpha|^2
+(\sigma_\alpha^2-2B_\al\sigma_\al)
|a_\alpha|^2\right]$ only in the absence of
dissipation ($\Gamma_\alpha=0$). We have checked that
angular momentum conservation ($J_s+J_{\rm orb}=$ const)
and ``energy conservation'' 
($E_{\rm trans}+E_{\rm orb}=$ const;
of course the energy of the system, $E_s+E_{\rm orb}$,
is not conserved due to dissipation)
are satisfied numerically to high accuracy.

Figure 7 reveals that, although initially 
the energy transfer during periastron passage varies from one orbit to
another, after a few dissipation time, the mode energy
$\langle E_\al\rangle$ averaged over an orbit reaches a 
constant. In this steady state, the energy transferred to a mode
at periastron is exactly balanced by the energy dissipated in one
orbital period. Moreover, the mean mode energy 
$\langle E_\al\rangle$ is approximately equal to the energy 
transfer in the first periastron passage, $\Delta E_\al$, 
which has been discussed in \S 3. This apparent coincidence
can be understood as follows. Consider a mode with a given $\al$
and sign of pattern speed. Its amplitude evolves according to
the first term of Eq.~(\ref{eqa1}), with the damping factor included:
\be
a_\al(t)={GM'W_{lm}Q_{\al}\over 2i(\sigma_\al-B_\al)}\,
e^{i\sigma_\al t-\Gamma_\al t}\int_{t_i}^t\!\!dt'\,
{e^{-im\Phi(t')-i\sigma_\al t'+\Gamma_\al t'}\over D^{l+1}(t')}.
\ee
Let $t_j=(2j-1)P_{\rm orb}/2$ (with $j=0,1,2,\cdots$) be the times at
apastron. After the $k$-th periastron passage, the mode amplitude
can be written as
\be
a_\al(t_k)=(\Delta a_\al) e^{(i\sigma_\al-\Gamma_\al)\,t_k}
\sum_{j=0}^{k-1}e^{(-i\sigma_\al+\Gamma_\al)j\Porb}
={e^{(i\sigma_\al-\Gamma_\al)\Porb/2}\over
1-e^{(i\sigma_\al-\Gamma_\al)\Porb}}
\left[1-e^{(i\sigma_\al-\Gamma_\al)k\Porb}\right](\Delta a_\al),
\ee
where $\Delta a_\al$ is the change of mode amplitude in the first
periastron passage when there is no initial oscillation
[cf.~Eq.~(\ref{eqa2})]
\be
\Delta a_\al={GM'W_{lm}Q_{\al}\over 2i(\sigma_\al-B_\al)}\,
\int_{-\Porb/2}^{\Porb/2}\!\!dt\,
{e^{-im\Phi-i\sigma_\al t}\over D^{l+1}}
={\pi GM'Q_\al\over iD_p^{l+1}(\sigma_\al-B_\al)}K_{lm}(\sigma_\al).
\ee
Clearly, when $k\Porb\Gamma_\al>>1$, the amplitude $a_\al(t_k)$
becomes independent of $k$. The
steady-state mode energy is given by
\be
\langle E_\al\rangle={\Delta E_\al\over 2\left[\cosh(\Gamma_\al\Porb)
-\cos(\sigma_\al\Porb)\right]}
\simeq {\Delta E_\al\over 4\sin^2(\sigma_\al\Porb/2)
+(\Gamma_\al\Porb)^2},
\label{esteady}\ee
where $\Delta E_\al=(GM'^2R^5/D_p^6)
\pi^2 |Q_\al K_{lm}(\sigma_\al)|^2\sigma_\al/(\sigma_\al-B_\al)$
is the energy transferred to the mode in the first periastron.
Equation (\ref{esteady}) establishes the relationship between
steady-state tidal energy $\langle E_s\rangle=\sum_\al\langle
E_\al\rangle$ and the initial energy transfer $\Delta
E_s=\sum_\al\Delta E_\al$. Typically we have 
$\langle E_s\rangle\simeq \Delta E_s$, 
except when a mode is in close resonance with the orbit,
i.e., $\sigma_\al=2\pi N/\Porb=N\Omega_{\rm orb}$ (where $N$ is an
integer). 

\subsection{Orbital Resonances}

We now consider the probability that the orbital evolution 
is driven by an near orbital resonance
$\sigma_\al\Porb=2(N+\eps)\pi$, with $|\eps|<<1$. For 
the modes which contribute significantly to the energy transfer,
$\sigma_\al\go 2\Omega_p$, thus
$N=\sigma_\al/\Omega_{\rm orb}\go 2(1+e)^{1/2}/(1-e)^{3/2}\simeq 32$.
For $\langle E_\al\rangle\ge 10\Delta E_\al$ we require
$|\eps|\le 0.05$, and we use this to define the full width
$(\Delta\Porb)_\eps$ (in the $\Porb$-space) of the resonance:
$(\Delta\Porb/\Porb)_\eps=2\eps/N$. Suppose 
the energy transfer is dominated by a single mode $\al$, 
then the time the orbit
spends near resonance is $(\Delta t)_\eps=(8\pi^2\eps^3/3N)
t_{\rm decay}$, where $t_{\rm decay}=|\Porb/\dot\Porb|$ with
$\dot\Porb$ evaluated assuming $\langle E_s\rangle\simeq
\langle E_\al\rangle=\Delta E_\al$. On the other hand, the times
it takes the orbit to evolve between resonances is
$(\Delta t)_r=(2/N)t_{\rm decay}$. Clearly, the ratio
$(\Delta t)_\eps/(\Delta t)_r=(4\pi^2/3)\eps^3$ is much less than 
unity. Therefore it unlikely that the orbital decay is driven by 
such a close resonance (see also KQ).
In the following, we shall set 
$\langle E_\al\rangle\simeq \Delta E_\al$.

\subsection{Retrograde vs. Prograde Rotations}

Figure 8 shows the dimensionless quantities [cf.~Eqs.~(\ref{eqdele2})
and (\ref{eqdelj2})] $T_2^{(r)}=T_{22}^{(r)}$
(energy transfer in the rotating frame), $S_2=S_{22}$ (angular
momentum transfer) and $T_2=T_{22}$ (energy transfer in 
the inertial frame) as a function of the stellar rotation rate
(for $\eta=7$, corresponding to neutron star mass $M'=1.4$ and
orbital semimarjor axis $20R$).
All g-modes with $n=1,\,2,\cdots,\,20$ are included 
in our calculations. As in prabolic tidal encounters (\S 5), 
we find that retrograde rotation increases the tidal
energy significantly because energy transfer is shifted
from higher-order to lower-order g-modes. 
We can see this effect clearly in Fig.~9, which 
shows contributions of different modes to the energy
transfer for several different rotation rates:
At $\Omega_s=0$, the dominant modes 
are g$_6$-g$_8$, while for $\Omega_s=-0.4$,
they are shifted to g$_3$-g$_5$. As a result, $\Delta E_s$ 
for $\Omega_s=-0.4$ is $24$ times larger than the nonrotating value.

We see from Fig.~8 that moderate prograde rotation $(\Omega_s>0)$
decreases $\Delta J_s$ and $\Delta E_s$,  
and for sufficiently large $\Omega_s$,
both $\Delta J_s$ and $\Delta E_s$ can become negative, i.e.,
angular momentum and energy are transferred from the star to the 
orbit, leading to orbital expansion. We find that $\Delta J_s<0$
for $\Omega_s>0.319$ and $\Delta E_s<0$ for $\Omega_s>0.362$. 
These critical $\Omega_s$'s are larger than $\Omega_p=0.192$, 
the orbital frequency at periastron
\footnote{The measured surface velocity of the B-star in the PSR
J0045-7319 binary is $113$ km/s (Bell et al.~1995). This corresponds
to a rotational angular frequency
[in units of $(GM/R^3)^{1/2}$] of $0.22/\sin i_{sn}$ (where $i_{sn}$
is the angle between the spin axis and the line-of-sight). 
The component of $\bOmega_s$ perpendicular to the orbital plane is
$(0.22/\sin i_{sn})\cos\beta$, where the 
spin-orbit inclination angle $\beta$ is constrained to be in the range
$25^o<\beta<55^o$ or $125^o<\beta<155^o$ (Kapsi et al.~1996;
note that the other constraint, $\beta<41^o$ or $\beta>139^o$,
is effective only if we restrict the orbital precession phase angle
to the first or the third quadrants).}.
Although this result
agrees qualitatively with that of KQ, there are significant
quantitative differences: KQ found that $\Delta E_s$ changes
sign at $\Omega_s/\Omega_p=1.2$ (from their Fig.~1 and and note that 
they used stellar radius $R=6R_\odot$) whereas we find
$\Omega_s/\Omega_p=1.9$ 
\footnote{This is different from static tide where $\dot E_{\rm orb}$
and $\dot J_{\rm orb}$ change sign at $\Omega_s\simeq \Omega_p$:
The weak friction theory (Alexander 1973; Hut 1981) gives
$\dot E_{\rm orb}\propto -[1-f(e)\Omega_s/\Omega_p]$
and $\dot J_{\rm orb}\propto -[1-g(e)\Omega_s/\Omega_p]$,
where $f=1,\,0.979,\,0.957$ and $g=1,\,1.216,\,1.212$
for $e=0,\,0.808,\,1$.}.
Moreover, KQ found that at extreme
rotation rate (say $\Omega_s>0.4$) the magnitude of the energy
transfer is larger than the nonrotating value (and they have
concluded that prograde rotation can also explain the observed 
orbital evolution timescale, although with a wrong sign), 
whereas we find that $|\Delta E_s|$ is much less than 
the nonrotating value even at $\Omega_s=0.5$.
We think these differences mainly come from KQ's 
problematic treatment of mode decomposition (particularly the
effect of Coriolis force). This is explained in the
following paragraphs.

KQ's results are based on
the solution of an approximate mode amplitude equation in the rotating
frame:
\be
\ddot a_\alpha+\omega_\alpha^2a_\alpha
+2\Gamma_\alpha\dot a_\alpha
={GM'W_{lm}Q_{\alpha}\over D^{l+1}}e^{-im(\Phi-\Omega_s t)},
\label{kqa}\ee
(only $l=2$ tides are considered). 
While they included the corrections to $\omega_\al$ and $Q_\al$ from the
Coriolis force, KQ neglected the corresponding 
Coriolis force term, proportional to $(\dot a_\al-i\omega_\al a_\al)$,
in Eq.~(\ref{kqa}).
Decomposing the external forcing $e^{-im\Phi}/D^{l+1}$ into Fourier
components with frequencies equal to multiples of $\Omega_{\rm orb}$,
we find that the steady-state mode amplitude is given by
\be
a_\alpha = {GM'Q_\alpha\over D_p^{l+1}}
\exp(im\Omega_s t)\sum_{k=-\infty}^\infty
{K_{lm,k}\,\exp(ik\Omega_{\rm orb} t)\over
\omega_\al^2-(m\Omega_s+k\Omega_{\rm orb})^2
+2\Gamma_\al(m\Omega_s+k\Omega_{\rm orb})i}
\label{asteady}\ee
where $K_{lm,k}\equiv \Omega_{\rm orb}K_{lm}(k\Omega_{\rm orb})$.
The mode amplitude in the inertial frame is simply
$a_\al\,e^{-im\Omega_s t}$. Note that $K_{lm,k}$ is very small
when $m$ and $k$ have the same signs, thus for $m>0$ we can neglect 
the positive-$k$ terms in the sum, and for $m<0$ neglect the
negative-$k$ terms. Equation (\ref{asteady}) is then 
equivalent to Eq.~(1) of KQ (see also Kumar, Ao \& Quataert 1995).
However, it is easy to see that this
expression is problematic: According to Eq.~(\ref{asteady}),
even at the apocenter, the mode does not oscillate at its intrinsic
frequency ($\omega_\al$ in the rotating frame), and yet its amplitude
still depends on the intrinsic mode quantities $\omega_\al$ and
$Q_\al$ (which are dependent on the pattern speed of the mode). 

To see the problem more clearly, we can calculate the energy transfer
and angular momentum transfer in the ``first'' periastron passage. 
As shown in \S6.1, in the absence of close orbital
resonance (\S 6.2), these are approximately equal to the 
steady-state mode energy and angular momentum. 
Applying the same procedure of \S 3 to Eq.~(\ref{kqa}),
we obtain the dimensionless tidal energy and angular momentum:
\begin{eqnarray}
T_2^{(r)}&=&\!\pi^2\sum_\al Q_\al^2\left[|K_{2m}(\omega_\al-m\Omega_s)|^2
+|K_{2-m}(\omega_\al+m\Omega_s)|^2\right],\label{kqtr}\\
S_2 &=&\!\pi^2\sum_\al Q_\al^2\left(-{m\over\omega_\al}\right)
\left[|K_{2m}(\omega_\al-m\Omega_s)|^2
-|K_{2-m}(\omega_\al+m\Omega_s)|^2\right],\label{kqs}\\
T_2 &=&\!\pi^2\sum_\al {Q_\al^2\over\omega_\al}\left[(\omega_\al-m\Omega_s)
|K_{2m}(\omega_\al-m\Omega_s)|^2
+(\omega_\al+m\Omega_s)|K_{2-m}(\omega_\al+m\Omega_s)|^2\right].
\label{kqt}
\end{eqnarray}
It is easy to see that Eqs.~(\ref{kqs})-(\ref{kqt})
are identical to Eqs.~(\ref{eqs}) and (\ref{eqt}) with $B_\al
=-m\Omega_s$ (i.e., with the Coriolis term neglected). 
For comparison, we also plotted in Fig.~8 the 
numerical results calculated using these expressions,
which are in reasonable agreement with KQ's (see their Fig.~1). 
However, these expressions are certainly {\it incorrect}, 
as they would imply that a prograde mode
and retrograde mode have the same frequencies in the rotating
frame. Indeed, integrating Eq.~(\ref{kqa}) directly
we find that after the first peristron passage, 
the mode amplitude (in the rotating frame) is a superposition 
of $e^{im\phi+i\omega_\al t}$ and $e^{im\phi-i\omega_\al t}$
--- This is clearly unphysical. It is easy to show that 
the errorous expressions (\ref{kqtr})-(\ref{kqt}) lead 
to overestimates of the negative angular momentum and energy
transfers for prograde stellar rotation.

\subsection{Long-Term Orbital Evolution}

In the steady state, the energy transferred to a stellar mode
in each periastron passage is equal to the mode energy dissipated 
in one orbit, $2\Gamma_\al\Porb\langle E_\al\rangle\simeq
2\Gamma_\al\Porb\Delta E_\al$. The corresponding
angular momentum transfer is 
$2\Gamma_\al\Porb\Delta J_\al$. Thus the orbital energy and angular
momentum decay rates are given by
\begin{eqnarray}
{d\langle E_{\rm orb}\rangle\over dt} &=& -\sum_\al
2\Gamma_\al \langle E_\al\rangle\simeq -\sum_\al 2\Gamma_\al
\Delta E_\al,\label{dedt}\\
{d\langle J_{\rm orb}\rangle\over dt} &\simeq& -\sum_\al
2\Gamma_\al \Delta J_\al=-{d\langle J_s\rangle\over dt}.
\label{djdt}
\end{eqnarray}
If we define the mean damping time $t_{\rm damp}$ via
$\sum_\al\Gamma_\al\Delta E_\al=\Delta E_s/t_{\rm damp}$,
then the current orbital decay rate is given by
\be
{\dot\Porb\over\Porb}\simeq -(0.5\,{\rm Myr})^{-1}
\left({T_2\over 10^{-2}}\right)\left({30\,{\rm yr}\over 
t_{\rm damp}}\right).
\ee
Note that $T_2=10^{-2}$ corresponds to nearly maximum retrograde
rotation. Thus to explain the observed $\dot\Porb$, the damping times
of the dominant modes (g$_3$-g$_5$; cf.~Fig.~9) must be less than
$30$ years. As pointed out by KQ, such a short radiative damping time
is unlikely if the star is rigidly rotating, but would be possible
if there is differential rotation. Such differential rotation is
naturally expected since tidal torque
deposits angular momentum mainly in the region near the stellar
surface (Goldreich \& Nicholson 1989). 

Independent of the mode damping rates, 
equations (\ref{dedt}) and (\ref{djdt}) provide a 
scaling relation for the orbital decay rate as 
a function of the orbital and spin parameters.
For $4<\eta<10$, we can fit the function $T_2$ by
$T_2\propto\eta^{-4\nu}$, where $\nu$ depends mainly on $\Omega_s$ but
only slightly on the eccentricity (as long as $e\go 0.7$): we find 
$\nu=1,\,0.5,\,0.2$ for $\Omega_s=0,\,-0.2,\,-0.4$ (see Fig.~4).
We then have
\be
{\dot\Porb\over\Porb}\propto -\Porb^{-10/3-4\nu}(1-e)^{-6(1+\nu)},
\ee
where the proportional constant can be fixed by the current
observed value of $\dot\Porb$. 
Similar equations for $\dot e$ and $\dot\Omega_s$ are easily 
obtained. 
These equations can be used to infer the long-term 
evolution of the binary system. Note that since $\Delta E_s
\simeq\beta(GM/R^3)^{1/2}\Delta J_s$, where $\beta$ is of order unity
(for the current system, $\eta=7$, we find $\beta=3$ for 
$\Omega_s=0$ and $\beta=5$ for $\Omega_s=-0.5$), 
the rates at which $J_{\rm orb}$
and $J_s$ change are given by
\be
\left|{\dot J_{\rm orb}\over J_{\rm orb}}\right|
={\beta\over 3}\left({1-e\over 1+e}\right)\Omega_p\left|{\dot\Porb
\over\Porb}\right|,~~~~~
\left|{\dot J_s\over J_s}\right|={0.50\,\beta(1-e)\over
\lambda (1+e)^{1/3}}\,\Omega_p^{2/3}\Omega_s^{-1}\left|{\dot\Porb
\over\Porb}\right|,
\label{djj}\ee
where we have taken the momont of inertia $I=0.1\lambda MR^2$,
and $\Omega_s$ and $\Omega_p$ are in units of $(GM/R^3)^{1/2}$.
Both $|\dot J_{\rm orb}/J_{\rm orb}|$ and $|\dot J_s/J_s|$ 
are much smaller than $|\dot\Porb/\Porb|$, i.e., the
timescale for changing the orbital angular momentum
and the timescale for synchonizing and aligning the stellar
spin are much longer than the orbital decay time
\footnote{This is different from evolution driven by static tide,
for which $\dot E_{\rm orb}\simeq \Omega_p\dot J_{\rm orb}$, and 
we can show that in this case $|\dot J_{\rm orb}/J_{\rm orb}|$ and
$|\dot J_s/J_s|$ are larger than those given in Eq.~(\ref{djj}) by 
a factor $\Omega_p^{-1}$ (although both are still less than 
$|\dot\Porb/\Porb|$).}.
The orbit therefore evolves with decreasing orbital
period and eccentricity, while
the periastron distance $D_p$ remains approximately constant.
Since $\Delta E_s$ mainly depends on $D_p$, the orbital decay rate
$|\dot\Porb/\Porb|\propto \Delta E_s/|E_{\rm orb}|$ was larger 
at earlier times (when $|E_{\rm orb}|$ was smaller). 
This implies that there is an upper limit to the age of the binary
since the neutron star formation. 
Integrating the parametrized evolutionary equations backward in time,
we find that regardless of the various uncertainties, the orbital age is 
always less than $1.4$ Myr (Lai 1996). 
This is significantly
smaller than the characteristic age ($3$ Myr) of the pulsar, 
implying that the latter is not a good age indicator.
The most likely explanation for this discrepancy is that
the initial spin period of the pulsar is close to its
current value. Thus the pulsar was either formed rotating very 
slowly, or has suffered spin-down due to accretion in the first $\sim
10^4$ years (the Kelvin-Holmholtz time of the B-star) after the
supernova (E.~van den Heuvel 1996, private communication).

\section{Discussion}

The main result of this paper has been summarized in the
abstract. Here we simply note that our study of dynamical tides in
rotating stars is based on an approximate scheme of mode
decomposition, which we believe gives physically meaningful
results. Nevertheless, a mathematically rigorous theory describing the 
response of individul modes in rotating stars
to external forcing is highly desirable, and the  
role of the continuous mode spectrum (Dyson \& Schutz 1979) need
to be clarified. 

In addition to the two applications (tidal capture binary formation
and the orbital evolution of pulsar binaries)
examined in this paper, tidal excitation of oscillation modes
may also be important in the final phase of inspiraling neutron star
binaries. Extracting gravitational wave signals from noise 
requires accurate theoretical 
waveforms in the frequency range $10-1000$ Hz, corresponding to the
last few minutes of the binaries' life (Cutler et al.~1993).
For nonrotating neutron star, 
the orbital phase error induced by resonant excitaion of g-mode
is negligible (Reisenegger \& Goldreich 1994; Lai 1994). 
With rapid rotation, the f-mode frequency 
in the inertial frame can be significantly reduced,  
therefore resonant excitation of f-mode is possible even 
at large orbital separation. We wish to study this and
related issues in a future paper.

\acknowledgments

I thank Pawan Kumar for sending me his papers prior to publication
and several interesting conversations.
I also thank Lars Bildsten, Lee Linblom and Greg 
Ushomirsky for discussing modes in rotating stars. 
This research is supported by the 
Richard C. Tolman Fellowship at Caltech, NASA Grant NAG 5-2756, and
NSF Grant AST-9417371.

\appendix

\section{Equations for General Spin-Orbit Inclination Angle}

In the main body of our paper we assume for simplicity that 
$\bOmega_s$ is aligned with the orbital angular momentum.
Generalization to the cases of arbitrary spin-orbit inclination
angle is straightforward.
Define a coordinate system $(xyz)$ centered on $M$ with 
the $z$-axis along the orbital angular momentum
and the $x$-axis pointing to the pericenter in the orbital plane. 
Define another coordinate system $(x'y'z')$ with 
the $z'$-axis along $\bOmega_s$ and the $y'$-axis in the orbital
plane. Let the angle between the $z$-axis and $z'$-axis be $\beta$
(the spin-orbit inclination angle), and the angle between 
the $y$-axis and $y'$-axis be $\alpha$. Thus the $(x'y'z')$ frame is
related to the $(xyz)$ frame by Euler angle $(\alpha,\beta,\gamma=0)$.
The function $Y_{lm}(\theta,\phi)$ is related to 
$Y_{lm'}(\theta',\phi')$ by
\be
Y_{lm}(\theta,\phi)=\sum_{m'}{\cal D}^{(l)}_{m'm}(\alpha,\beta)
Y_{lm'}(\theta',\phi'),
\ee
where the Wigner ${\cal D}$-function is given by
\begin{eqnarray}
{\cal D}^{(l)}_{m'm}(\alpha,\beta)&=&e^{im'\alpha}
\left[(l+m)!\,(l-m)!\,(l+m')!\,(l-m')!\right]^{1/2}\nonumber\\
&&\times \sum_k{(-1)^{l+m'-k}\left(\cos{1\over
2}\beta\right)^{2k-m-m'}\!\!
\left(\sin{1\over 2}\beta\right)^{2l-2k+m+m'}\over
k!\,(l+m-k)!\,(l+m'-k)!\,(k-m-m')!},
\end{eqnarray}
(e.g., Wybourne 1974). The mode amplitude equation (\ref{eqa})
should then be replaced by 
\be
\ddot a_\alpha+\sigma_\alpha^2a_\alpha
-2iB_\alpha(\dot a_\alpha-i\sigma_\alpha a_\alpha)
+2\gamma_\alpha\dot a_\alpha
=\sum_{lm'} {GM'W_{lm'}Q_{\alpha l}\over D^{l+1}}
{\cal D}^{(l)}_{mm'}e^{-im'\Phi}.
\ee
It is easy to show that in the egneral cases, equations 
(\ref{eqa2})-(\ref{eqs2}) still apply, provided that 
$K_{lm}(\sigma_\al)$ is replaced by 
\be
{\tilde K}_{lm}(\sigma_\al)=\sum_{m'}
{\cal D}^{(l)}_{mm'}K_{lm'}(\sigma_\al),
\ee
[recall that $\al=\{njm\}$]. Note that in general,
the $m=\pm 1$ modes also contribute to the tidal excitation
(unlike the aligned case, where only the $m=0,\,\pm 2$ modes
are excited).

\section{The Functions $Q_{\alpha l}$ and $B_\alpha$}

With $\xi_r(\br)$ and $\delta P(\br)$ expressed in Eq.~(\ref{eigen}), 
the other eigenfunctions are
\begin{eqnarray}
\delta\rho(\br) &=&\delta\rho(r)H e^{im\phi}
=\left({\delta P\over c_s^2}+{\rho N^2\over g}\xi_r\right)
H e^{im\phi},\\
\xi_\theta(\br) &=& {\xi_\perp(r)\over 1-q^2\mu^2}
\left({\partial H\over\partial\theta}
+{mq\mu\over\sin\theta}H\right)e^{im\phi},\\
\xi_\phi(\br) &=&{i\xi_\perp(r)\over 1-q^2\mu^2}
\left(q\mu{\partial H\over\partial\theta}
+{m\over\sin\theta}H\right)e^{im\phi},
\end{eqnarray}
where the mode index $\alpha=\{njm\}$ has been surpressed, 
$q=2\Omega_s/\omega_\alpha$, 
$\mu=\cos\theta$, and $\xi_\perp(r)=(\delta P)/(\rho r\omega^2)$.
We normalize the Hough function via
$\int\!d\Omega H^2(\theta)=2\pi\int\!d\mu
H^2=1$. The eigenfunction is normalized according to
\be
\int\!d^3x\,\rho\,\vxi^\ast\cdot\vxi=\int_0^Rdr\,r^2\rho\left(\xi_r^2
+\Lambda\xi_\perp^2\right)=1,
\ee
where
\be
\Lambda=2\pi\int_{-1}^1{d\mu\over (1-q^2\mu^2)^2}
\left[(1+q^2\mu^2)(1-\mu^2)
\left({\partial H\over\partial\mu}\right)^2
+{m^2(1+q^2\mu^2)\over 1-\mu^2}H^2-4mq\mu H
\left({\partial H\over\partial\mu}\right)\right].
\ee
Note that $\Lambda\rightarrow l(l+1)$ as $q\rightarrow 0$.
The tidal coupling coefficient defined by Eq.~(\ref{qalpha}) is
\be
Q_{\alpha l}=Q_\theta\int_0^R\!dr\,r^{l+2}\delta\rho_\alpha(r),~~~~
Q_\theta=\int\!d\Omega H e^{-im\phi}Y_{lm}.
\ee
The function $B_\alpha$ defined by Eq.~(\ref{balpha}) is
\begin{eqnarray}
B_\alpha&=&-m\Omega_s+2i\Omega_s\int\!d^3x\,\rho\,\xi_\phi^\ast
(\xi_\theta\cos\theta+\xi_r\sin\theta)\nonumber\\
&=&-m\Omega_s+2\Omega_s\left(B_{\theta 1}
\int_0^R\!dr\,r^2\rho\xi_\perp^2
+B_{\theta 2}\int_0^R\!dr\,r^2\rho\xi_r\xi_\perp\right),
\end{eqnarray}
where
\begin{eqnarray}
B_{\theta 1}&=&2\pi\int_{-1}^1\!{\mu d\mu\over (1-q^2\mu^2)^2}
\left[q\mu(1-\mu^2)\left({\partial H\over\partial\mu}\right)^2
+{m^2q\mu\over 1-\mu^2}H^2-m(1+q^2\mu^2)H{\partial H\over\partial\mu}
\right],\\
B_{\theta 2}&=&2\pi\int_{-1}^1\!{d\mu \over (1-q^2\mu^2)}
\left[mH^2-q\mu(1-\mu^2)H{\partial H\over\partial\mu}\right].
\end{eqnarray}
Note that $B_{\theta 1}\rightarrow m/2$ and $B_{\theta 1}\rightarrow m$
as $q\rightarrow 0$. 

\section{Asymptotic Expressions for $K_{lm}(\omega)$}

For parabolic orbit and in the limit $y\equiv\eta\,\sigma>>1$ 
(where $\sigma=\omega-m\Omega_s$ is the mode frequency
in the inertial frame), the functions $K_{lm}(\sigma)$ can be
evaluated analytically. Following PT, we write
\be
K_{lm}(\omega)={\sqrt{2}W_{lm}\over\pi}\,\eta\,I_{lm}(y),
\ee
\be
I_{lm}(y)=\int_0^\infty\!dx(1+x^2)^{-l}\cos\left[2^{1/2}y(x+{x^3/3})
+2m\tan^{-1}x\right].
\label{Ilm}\ee
Using the method of steepest decents, the integral (\ref{Ilm})
can be evaluated in the asymptotic limit ($y>>1$). For $l=2$, the
leading terms are
\begin{eqnarray}
I_{2-2}(y)&=&{2\pi^{1/2}\over
3}z^{3/2}\exp(-2z/3)\left(1-{\pi^{1/2}\over 4}z^{-1/2}+\cdots\right),\\
I_{20}(y)&=&{\pi^{1/2}\over 4}z^{1/2}\exp(-2z/3)
\left(1+{\pi^{1/2}\over 2}z^{-1/2}+\cdots\right),\\
I_{22}(y)&=&{\pi^{1/2}\over 32}z^{-1/2}\exp(-2z/3)
\left(1-{89\over 48}z^{-1}+\cdots\right),
\end{eqnarray}
where $z=\sqrt{2}\,y$ and we have assumed $y>0$.
Comparison with numerical integration indicates that
these asymptotic expressions are accurate to within $2\%$ 
for $y>2$.


\begin{table}
\caption{Fitting Parameters for $T_2$}
\vspace{0.3cm}
\begin{tabular}{ccc}
\hline 
\hline 
$~~~~~~\Omega_s~~~~~~$& $~~~~~~A~~~~~~$ & $~~~~~~\alpha~~~~~~$\\ 
\hline 
\multicolumn{3}{l}{$\Gamma=4/3$ Model:
$~~~T_2=A\,\eta^{-\alpha}$}\\
\hline 
-0.5 &0.25&1.80\\
-0.4 &0.25&2.00\\
-0.3 &0.25&2.26\\
-0.2 &0.25&2.51\\ 
-0.1 &0.25&2.80\\
0 &0.24&3.1\\
0.1 &0.27&3.6\\
0.2 &0.28&4.1\\
0.3 &0.29&4.6\\
0.4 &0.31&5.2\\
0.5 &0.33&5.8\\
\hline 
\multicolumn{3}{l}{$\Gamma=5/3$ Model: $~~~T_2=A\,\eta^5
\exp({-\alpha\eta})$}\\
\hline 
-0.6 &0.51&1.57\\
-0.4 &1.3&1.94\\
-0.2 &3.4&2.34\\
0 &8.1&2.74\\ 
0.2 &14.&3.10\\
0.4 &24.&3.45\\
0.6 &41.&3.80\\
\hline 
\hline 
\end{tabular} 
\end{table}
\clearpage

\bigskip
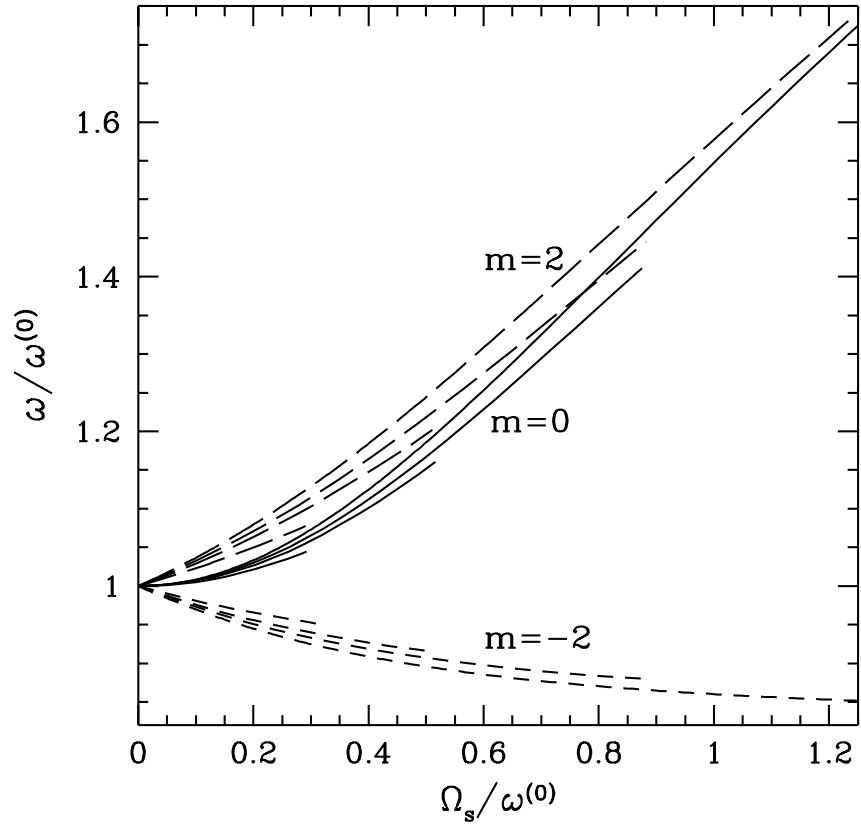
\noindent {\bf Figure 1} -- 
Frequencies $\omega$ (in the rotating frame) 
of selected $j=2$ (correcponding to $l=2$ in the
$\Omega_s\rightarrow 0$ limit) g-modes 
as a function of rotation rate $\Omega_s$. Both
$\omega$ and $\Omega_s$
are expressed in units of $\omega^{(0)}$, the corresponding mode frequency
at zero rotation. The solid curves are for $m=0$, short-dashed curves 
for $m=-2$, and long-dashed curves for  $m=2$. For each $m$, 
the four curves correspond g-modes of different radial order:
g$_2$, g$_5$, g$_{10}$ [these curves terminate at
$\Omega_s=0.5$], and g$_\infty$ (the WKB limit). The
numerical results shown in this figure 
can be fitted by Eqs.~(\ref{eqm2})-(\ref{eqm-2}).

\bigskip
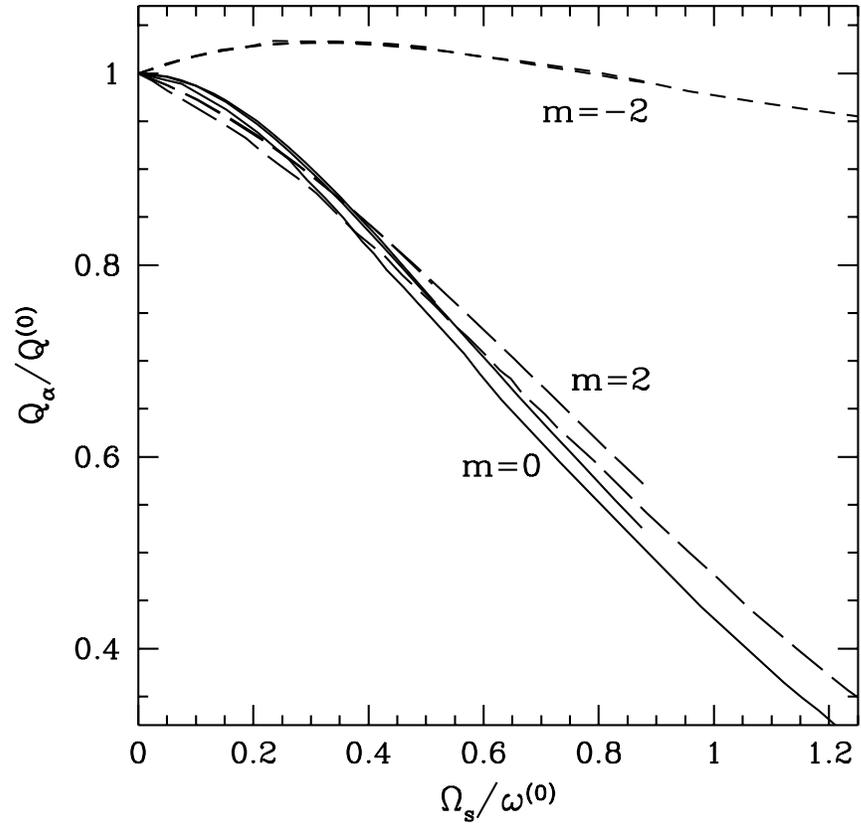
\noindent {\bf Figure 2} -- 
The tidal coupling coefficients 
$Q_\al=Q_{\alpha 2}$ of selected $j=2$ modes
as a function of rotation rate $\Omega_s$ (in units of
$\omega^{(0)}$, the corresponding mode frequency at zero rotation).
The solid curves are for $m=0$, short-dashed curves 
for $m=-2$, and long-dashed curves for $m=2$. For each $m$, 
the three curves correspond g-modes of different radial order:
g$_5$, g$_{10}$ (these curves terminate at $\Omega_s=0.5$)
and g$_{20}$ (Note that the g$_5$ curves almost overlap with the
g$_{10}$ curves).  

\bigskip
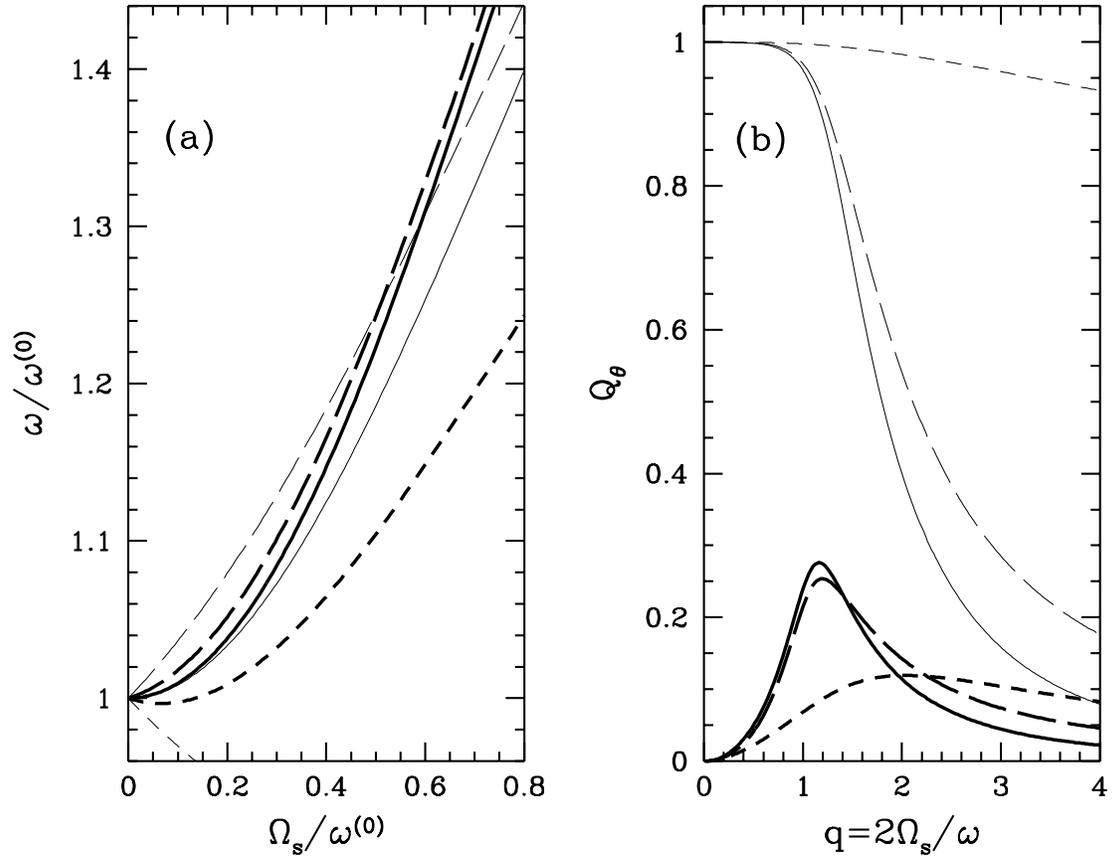
\noindent {\bf Figure 3} -- 
The frequencies $\omega$ and angular coupling coefficients $Q_\theta$
of selected $j=4$ g-modes (heavy curves) compared with those
of the $j=2$ modes (lighter curves). The solid lines are for $m=0$,
short-dashed lines for $m=-2$ and long-dashed lines for $m=2$. 
(a) gives the frequency (in the rotating frame)
$\omega$ in the WKB limit ($n\rightarrow\infty$),
and (b) shows the quantity $Q_\theta$ as defined in Appendix B. 

\bigskip
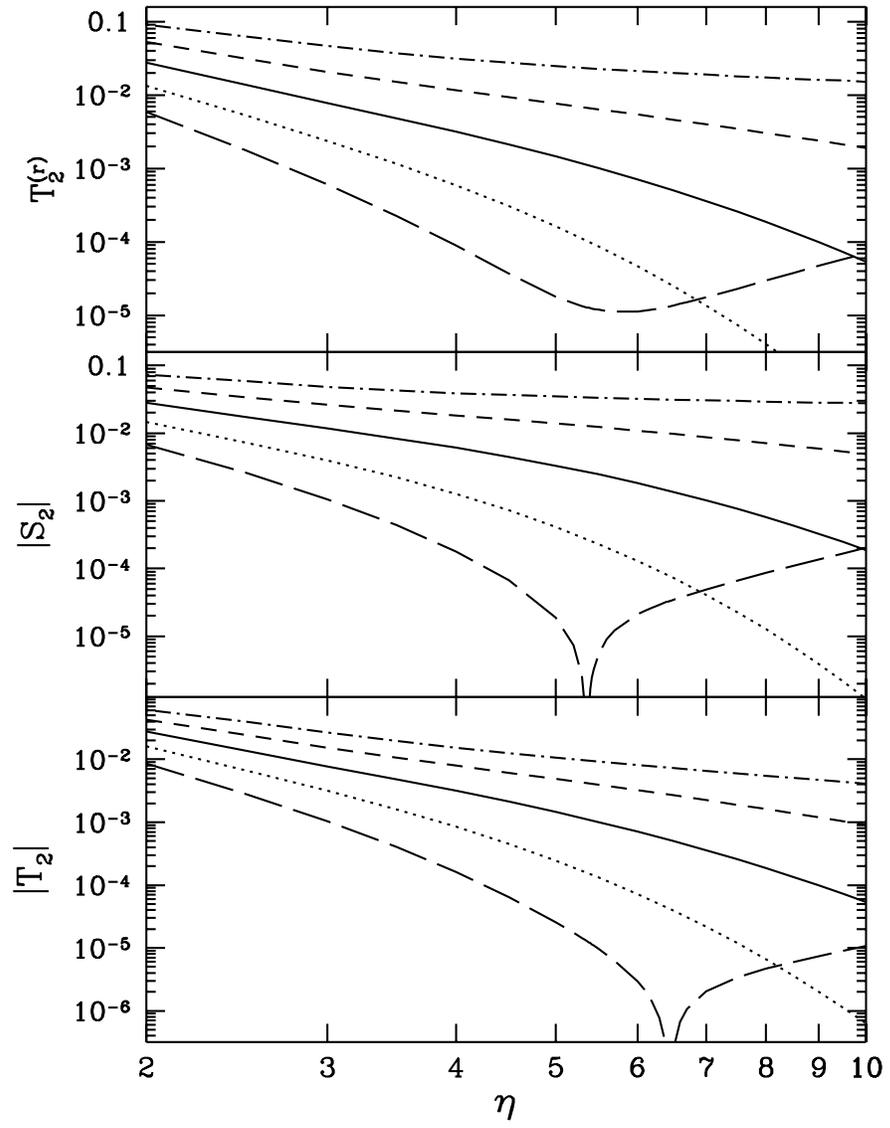
\noindent {\bf Figure 4} -- 
Dimensionless functions $T_2^{(r)},~S_2$ and
$T_2=T_2^{(r)}+\Omega_sS_2$ [where
$\Omega_s$ is the rotation rate in units of $(GM/R^3)^{1/2}$]
as a function of $\eta$ for $\Omega_s=0$ (solid lines),
$0.2$ (dotted lines), $-0.2$ (short-dashed lines),
$0.4$ (long-dashed lines) and $-0.4$ (dot-dashed lines)
in a parabolic encounter. The star is modeled as a $\Gamma=4/3$,
$\Gamma_1=5/3$ polytrope. Note that for $\Omega_s=0.4$ (long-dashed
lines), $S_2$ becomes negative for $\eta>5.36$ [see (b)], and
$T_2$ becomes negative for $\eta>6.45$ [see (c)].

\bigskip
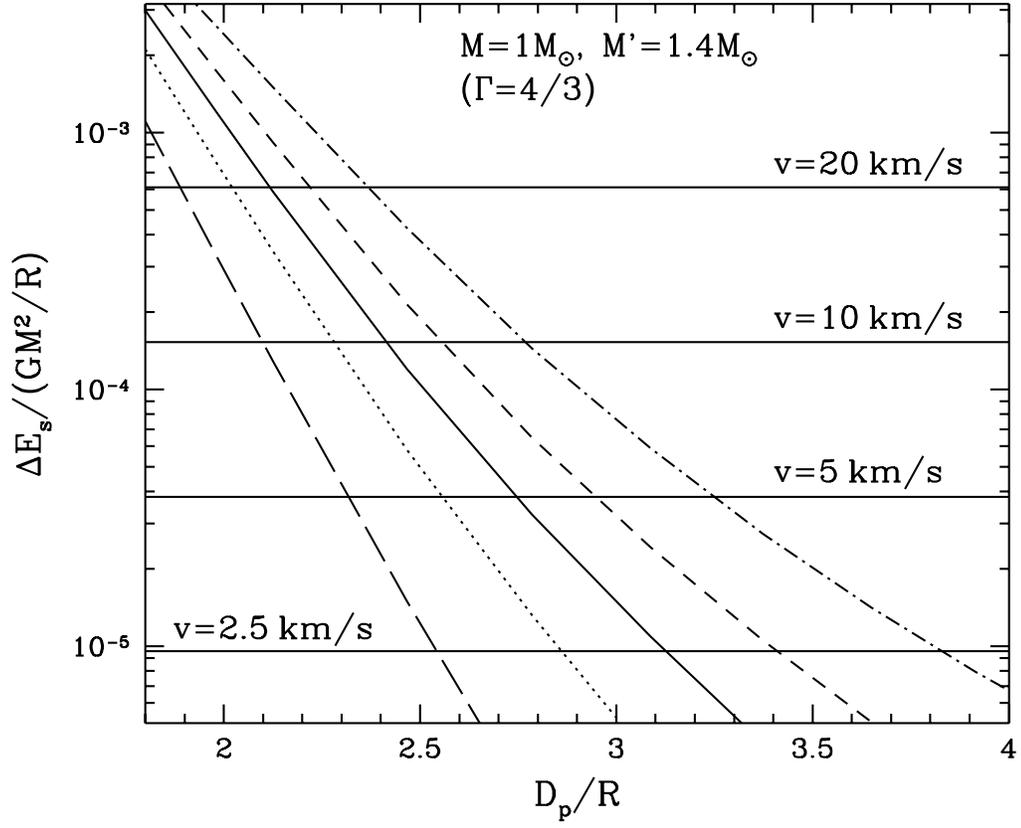
\noindent {\bf Figure 5} --
Energy transfer as a function of periastron distance $D_p$
during an encounter between
a main-sequence star $M=1M\odot$, $R=1R_\odot$ (modelled
as a $\Gamma=4/3$ polytrope) and a $M'=1.4M_\odot$ point mass (neutron
star). The four horizontal lines show the kinetic energy at infinity
$\mu v^2/2$ of the two stars with relative velocity 
$v=20,~10,~5,~2.5$ km$\,$s$^{-1}$. The solid curve is for 
$\Omega_s=0$, the dotted curve $\Omega_s=0.2$, the long-dashed
curve $\Omega_s=0.5$ (close to the maximum rotation rate),
the short-dashed curve $\Omega_s=-0.2$,
and the dot-dashed curve $\Omega_s=-0.5$.

\bigskip
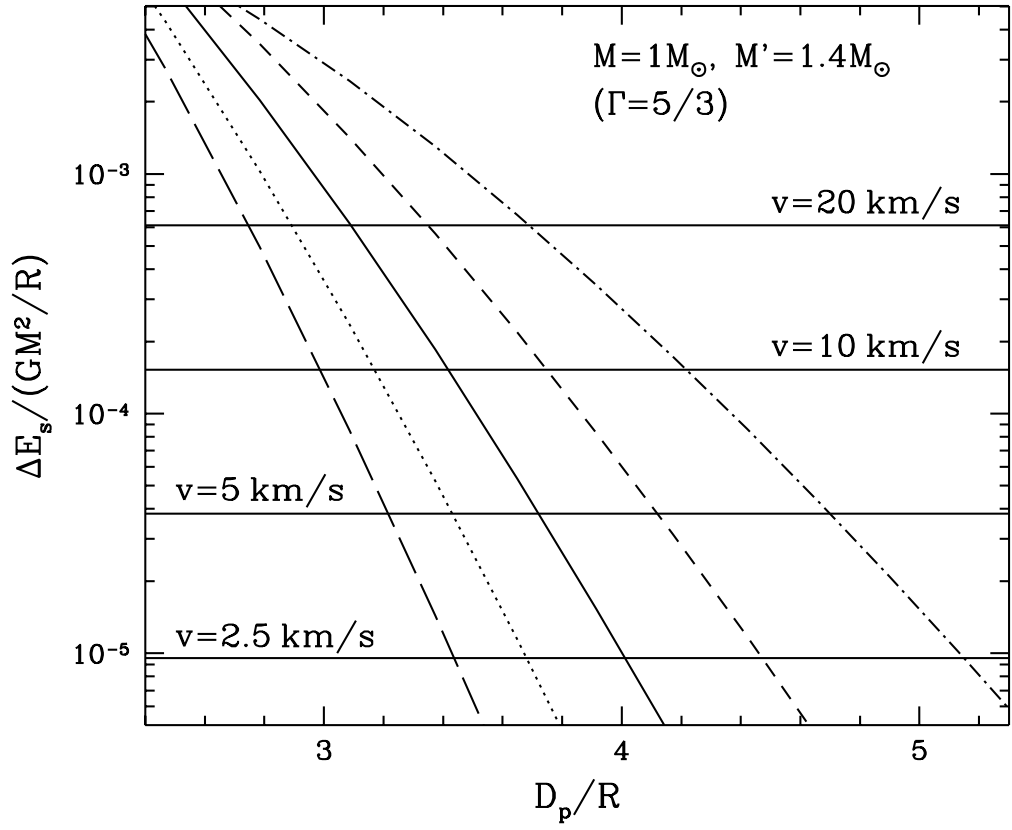
\noindent {\bf Figure 6} --
Energy transfer as a function of periastron distance $D_p$
during an encounter between
a main-sequence star $M=1M\odot$, $R=1R_\odot$ (modeled
as a $\Gamma=5/3$ polytrope) and a $M'=1.4M_\odot$
point mass (neutron star).
The four horizontal lines show the kinetic energy at infinity
$\mu v^2/2$ of the two stars with relative velocity 
$v=20,~10,~5,~2.5$ km$\,$s$^{-1}$. The solid curve is for 
$\Omega_s=0$, the dotted curve $\Omega_s=0.3$, the long-dashed
curve $\Omega_s=0.6$ (close to the maximum rotation rate), 
the short-dashed curve $\Omega_s=-0.3$,
and the dot-dashed curve $\Omega_s=-0.6$.

\bigskip
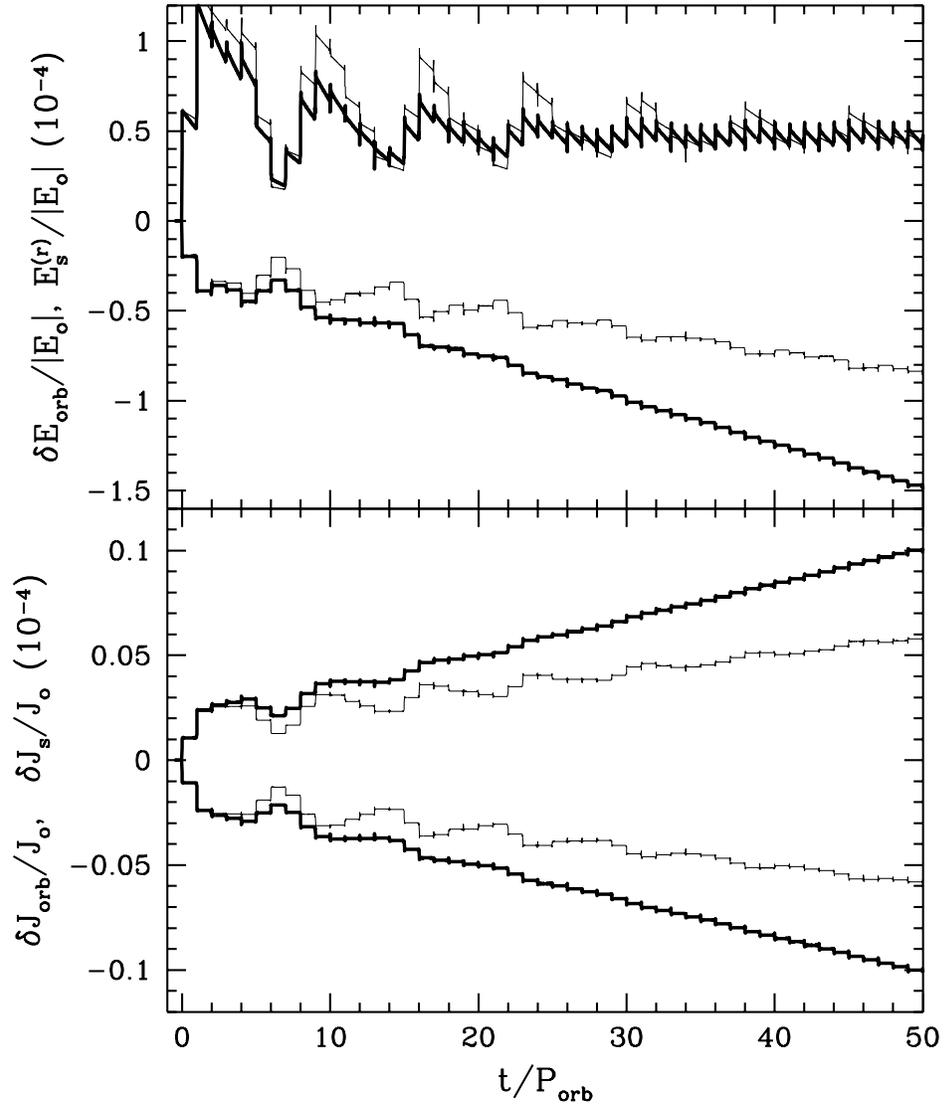
\noindent {\bf Figure 7} --
Evolution of (a) the energies $\delta E_{\rm orb}=E_{\rm orb}-E_o$
(where $E_o$ is the initial orbital energy) and $E_s^{(r)}$ (the mode
energy in the rotating frame), and (b) the angular momenta 
$\delta J_{\rm orb}=J_{\rm orb}-J_o$ (where $J_o$ is the initial
orbital angular momentum) and $\delta J_s$ (the mode angular 
momentum) due to dynamical tide with $\Omega_s=-0.4$. Only g$_2$-g$_8$
modes are included in the calculation, and relatively large damping
rates are chosen for clearer illustration: $\Gamma_\al=0.1/P_{\rm orb}$
(heavy lines) and $\Gamma_\al=0.05/P_{\rm orb}$ (light 
lines). 

\bigskip
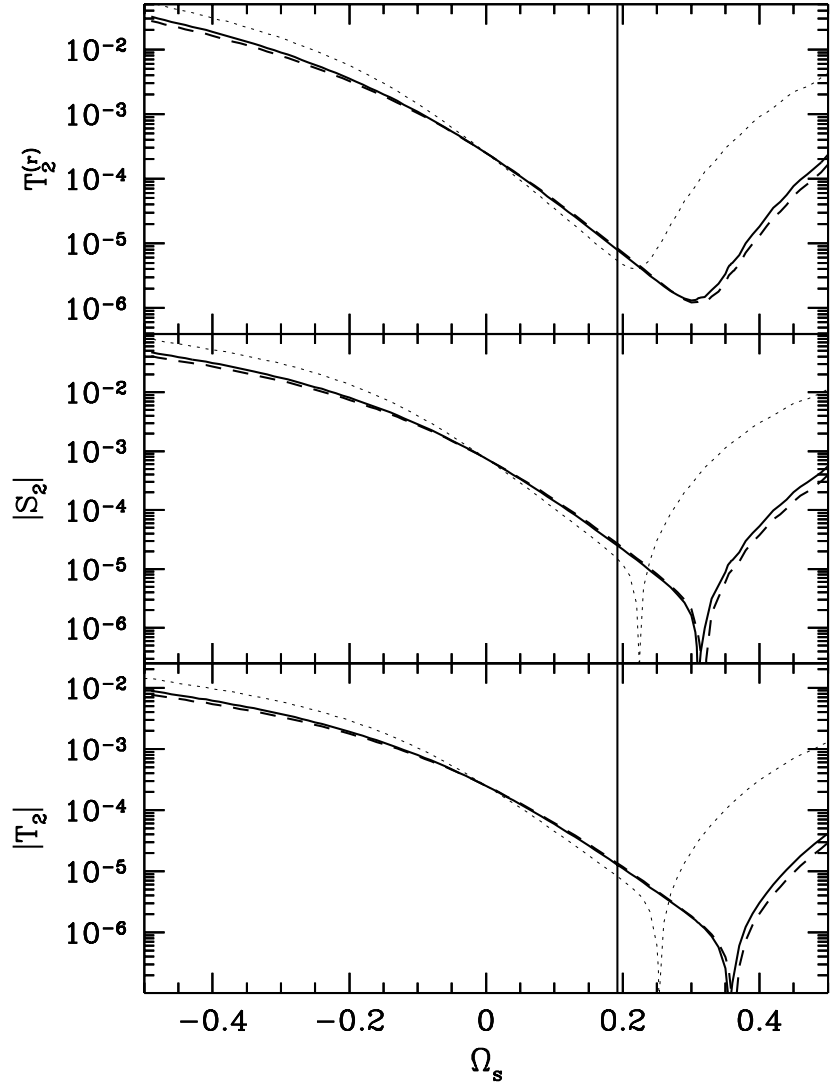
\noindent {\bf Figure 8} --
Dimensionless tidal energies $T_2^{(r)}$ (in the rotating frame),
$T_2$ (in the inertial frame) and angular momentum $S_2$
as a function of the stellar rotation rate $\Omega_s$ [in units
of $(GM/R^3)^{1/2}$]. The dimensionless orbital parameters are
$\eta=7$ and $e=0.808$. The solid curves are obtained 
using the correct expressions (\ref{eqt2})-(\ref{eqtr});
the dashed curves are obtained using the same equations
but assuming $B_\al=-m\Omega_s$ (i.e., neglecting the Coriolis 
term) --- These demonstrate that our results are insensitive 
to uncertainties in Eqs.~(\ref{eqt2})-(\ref{eqtr}). 
Note that $S_2$ and $T_2$ become negative for $\Omega_s>0.311$
and $\Omega_s>0.356$ respectively. 
The dotted curves are obtained using the incorrect
expressions (\ref{kqtr})-(\ref{kqt}); these give results
similar to those of KQ. 

\bigskip
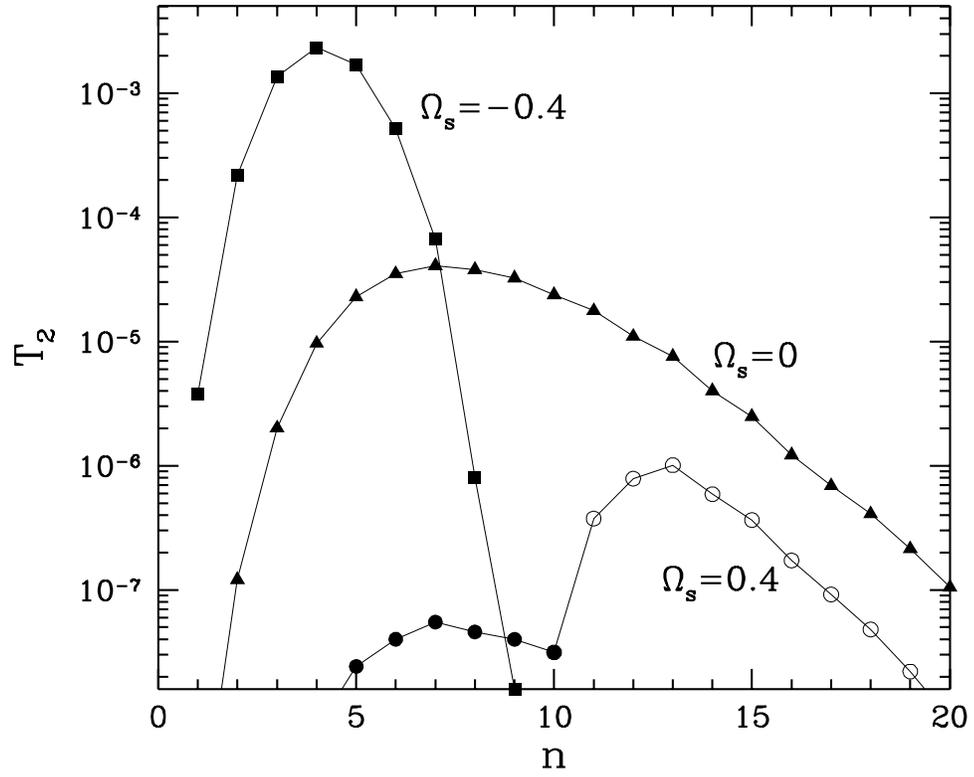
\noindent {\bf Figure 9} --
The contributions of g-modes of different order ($n$) 
to the function $T_2$ [the dimensionless energy transfer in the
inertial frame; cf.~Eqs.~(\ref{eqdele2}), (\ref{eqt2})]
for $\Omega_s=0$ (triangles), $\Omega_s=-0.4$ (squares)
and $\Omega_s=0.4$. The solids circles indicate positive
contributions, the open circles indicate negative ones.

\end{document}